\documentclass[reprint,nofootinbib,amsmath,amssymb,showkeys,aps]{revtex4-1}

\usepackage{graphicx}
\usepackage{bm}
\usepackage{natbib}
\usepackage{comment}
\usepackage[normalem]{ulem}

\usepackage[dvipsnames]{xcolor}
  
\usepackage[T1]{fontenc} % if needed
\usepackage{braket}

\definecolor{lcol}{HTML}{006699} %Luciano
\usepackage[nameinlink,noabbrev]{cleveref}
\Crefname{equation}{Eq.}{Eqs.}
\Crefname{section}{Sect.}{Sects.}
\Crefname{figure}{Fig.}{Figs.}

\defcitealias{IST:paper1}{EC19}

\begin{document}

\title{Early dark energy in the pre- and post-recombination epochs}

\author{Adri\`a G\'omez-Valent$^1$}\email{gomez-valent@thphys.uni-heidelberg.de}
\author{Ziyang Zheng$^1$}\email{zheng@thphys.uni-heidelberg.de}
\author{Luca Amendola$^1$}	\email{l.amendola@thphys.uni-heidelberg.de}
\author{Valeria Pettorino$^2$}\email{valeria.pettorino@cea.fr}
\author{Christof Wetterich$^1$}\email{c.wetterich@thphys.uni-heidelberg.de}

\affiliation{$^1$ Institut f\"{u}r Theoretische Physik, Ruprecht-Karls-Universit\"{a}t Heidelberg, Philosophenweg 16, D-69120 Heidelberg, Germany}
\affiliation{$^2$ AIM, CEA, CNRS, Universit{\'e} Paris-Saclay, Universit{\'e} Paris Diderot, 
             Sorbonne Paris Cit{\'e}, F-91191 Gif-sur-Yvette, France}

\begin{abstract}
Dark energy could play a role at redshifts $z\gg\mathcal{O}(1)$. Many quintessence models possess scaling or attractor solutions where the fraction of dark energy follows the dominant component in previous epochs of the Universe's expansion, or phase transitions may happen close to the time of matter-radiation equality. A non-negligible early dark energy (EDE) fraction around matter-radiation equality could contribute to alleviate the well-known $H_0$ tension. In this work we constrain the fraction of EDE using two approaches: first, we use a fluid parameterization that mimics the plateaux of the dominant components in the past. An alternative tomographic approach constrains the EDE density in binned redshift intervals. The latter allows us to reconstruct the evolution of $\Omega_{de}(z)$ before and after the decoupling of the Cosmic Microwave Background (CMB) photons. We have employed {\it Planck} data 2018, the Pantheon compilation of supernovae of Type Ia (SNIa), data on galaxy clustering, the prior on the absolute magnitude of SNIa by SH0ES, and weak lensing data from KiDS+VIKING-450 and DES-Y1. When we use a minimal parameterization mimicking the background plateaux, EDE has only a small impact on current cosmological tensions. We show how the constraints on the EDE fraction weaken considerably when its sound speed is allowed to vary. By means of our binned analysis we put very tight constraints on the EDE fraction around the CMB decoupling time, $\lesssim 0.4\%$ at $2\sigma$ c.l. We confirm previous results that a significant EDE fraction in the radiation-dominated epoch loosens the $H_0$ tension, but tends to worsen the tension for $\sigma_8$. A subsequent presence of EDE in the matter-dominated era helps to alleviate this issue. When both the SH0ES prior and weak lensing data are considered in the fitting analysis in combination with data from CMB, SNIa and baryon acoustic oscillations, the EDE fractions are constrained to be $\lesssim 2.6\%$ in the radiation-dominated epoch and $\lesssim 1.5\%$ in the redshift range $z\in (100,1000)$ at $2\sigma$ c.l. The two tensions remain with a statistical significance of $\sim 2-3\sigma$ c.l.
\end{abstract}


\maketitle
%\tableofcontents

%%%%%%%%%%%%%%%%%%%%%%%%%%%%%%%%%%%%%%%%%%%%%%%%%%%%%%%%%%%%%%%%%
%%%%%%%%%%%%%%%%%%%%%%%%%%%%%%%%%%%%%%%%%%%%%%%%%%%%%%%%%%%%%%%%%
%%%%%%%%%%%%%%%%%%%%%%%%%%%%%%%%%%%%%%%%%%%%%%%%%%%%%%%%%%%%%%%%%

\section{Introduction}\label{intro}
Various tensions in cosmological observations could indicate that the minimal model for dark energy, namely the cosmological constant, may be insufficient. The observed present value of the Hubble parameter $H_0$ seems to be higher than the one inferred from the Cosmic Microwave Background (CMB) and other observations measuring the early state of the Universe \cite{Planck:2018vyg,Riess:2020fzl,Verde:2019ivm,DiValentino:2020zio}. 
On the other hand, the amount of large-scale structure (LSS) in the current Universe could be lower than the one inferred from early cosmology \cite{Planck:2018vyg,Joudaki:2019pmv,Wright:2020ppw,Heymans:2020gsg,DiValentino:2020vvd}. 
If these tensions grow with future data, and systematics are under control, a possibility is that
dark energy is dynamical \cite{Wetterich:1987fm,Peebles:1987ek,Ratra:1987rm} in one form or another.

Dynamical dark energy can be associated with a scalar field whose potential and kinetic energies evolve with time and are (almost) homogeneously distributed in the Universe. Such models of quintessence could help to understand why dark energy plays an important role in present cosmology without an extreme tuning of model parameters. The early cosmological evolution may obey a scaling behavior \cite{Wetterich:1987fm,Copeland:1997et} for which the dark energy density is proportional to the dominant energy density in radiation or matter. Due to its decrease with time the present matter density is very small in Planck units, and the same would hold for the dynamical dark energy density. Such models lead to a small fraction of ``early dark energy'' (EDE) for high redshift $z$, in contrast to a negligible value for the cosmological constant.

Some cosmological event or change of evolution properties has to occur in order to end such a scaling solution and induce the much slower evolution of the scalar field that can be inferred from the observed accelerated expansion of the Universe and present dark energy domination (this is for example the case of growing neutrino cosmologies, see \citet{Wetterich_2007, Amendola_2008, Pettorino_2010}). We treat here the dynamics of this end of the scaling solution as unknown and gather information on the amount of EDE at different cosmological epochs from observations in a model-independent way. Early dark energy arises also in many other types of models, as for example in ultra-light axion-like models \cite{Poulin:2018dzj} or in those linked to late phase transitions \cite{Niedermann:2019olb,Gogoi:2020qif,Niedermann:2020dwg}. Our analysis is thought to cover a large class of EDE-models.

Dynamical dark energy could be coupled to dark matter \cite{Wetterich:1994bg,Amendola:1999er}.
This modifies both the background evolution and the formation of structure. Many models of modified gravity are actually equivalent to some version of coupled dark energy \cite{Wetterich:2014bma, Pettorino_2008}. A nonzero coupling impacts indeed the tensions in cosmological observations \cite{2013PhRvD..88f3519P, Planck:2015bue} - for a recent update see \cite{Gomez-Valent:2020mqn} and references therein. In the present paper we investigate the case of uncoupled quintessence in order to find out if some appropriate time evolution of EDE could remove the tensions, as advocated, for example, in different versions of ``new EDE'' \cite{Poulin:2018cxd,Agrawal:2019lmo,Smith:2019ihp,Niedermann:2020dwg,Gogoi:2020qif,Niedermann:2019olb}.

The amount of early dark energy is, however, severely constrained by the combination of several cosmological observations \cite{Doran:2006kp,Calabrese:2011hg,Pettorino:2013ia,Hojjati:2013oya,Planck:2015bue,Karwal:2016vyq,Hill:2020osr,Ivanov:2020ril,Chudaykin:2020acu,Seto:2021xua,DAmico:2020ods}. These constraints depend on the detailed time evolution of EDE, see e.g. \citet{Pettorino:2013ia} for an estimate  within different parameterizations and different epochs at which EDE starts/ends to be relevant. The aim of the present paper is to continue this study with a more detailed understanding of the constraints on EDE for various different redshift ranges, and assessment of the possible role of EDE for the observed tensions in the light of these constraints. For this purpose we propose a new, more general parameterization that encompasses most models encountered so far, as well as a tomographic approach. We further investigate the impact of assumptions on the sound speed on constraining EDE. In our analysis we include the CMB temperature, polarization and lensing data from {\it Planck} 2018, the Pantheon compilation of supernovae of Type Ia (SNIa), data on galaxy clustering from several surveys, the prior on the absolute magnitude of SNIa obtained from the first steps of the cosmic distance ladder by SH0ES, and weak lensing data from KiDS+VIKING-450 and DES-Y1. This data set constitutes a significant
improvement with respect to previous analyses, e.g. \cite{Pettorino:2013ia,Planck:2015bue}.

%%%%%%%%%%%%%%%%%%%%%%%%%%%%%%%%%%%%%%%%%%%%%%%%%%%%%%%%%%%%%%%%%
%%%%%%%%%%%%%%%%%%%%%%%%%%%%%%%%%%%%%%%%%%%%%%%%%%%%%%%%%%%%%%%%%
%%%%%%%%%%%%%%%%%%%%%%%%%%%%%%%%%%%%%%%%%%%%%%%%%%%%%%%%%%%%%%%%%

\section{Early dark energy}\label{sec:EDE}

Dark energy models may have  different amounts of dark energy at different epochs, which distinguishes them from a $\Lambda$CDM model. Measuring the amount of EDE at different epochs in the past can then be the key to distinguish such scenarios and possibly falsify a cosmological constant. 
EDE affects observables in various ways. First, the presence of EDE at decoupling can change the position and height of the peaks in the CMB \cite{Doran:2000jt, Pettorino:2013ia,Planck:2015bue}: this is a consequence of the fact that EDE affects the distance to last scattering surface, as well as the relative amount of DE with respect to other components (such as dark matter or neutrinos). EDE can also impact the CMB anisotropies through the early integrated Sachs-Wolfe effect. Furthermore, EDE suppresses the growth of structure after last scattering \cite{Ferreira:1997au,Doran:2001rw,Caldwell:2003vp}: a smaller number of clusters can form with respect to a cosmological constant scenario \cite{Grossi_2009}; the lensing potential is also weaker, with an impact on both weak lensing and the smoothing of CMB peaks at large multipoles due to CMB lensing.  

It is then clear that a sufficient amount of EDE, if compatible with other observations, can potentially have a direct impact on both $H_0$ and $\sigma_8$ tensions. In the following we first design a general parameterization able to mimic the background dominant component, and then proceed with a tomographic analysis in different redshift bins.

%%%%%%%%%%%%%%%%%%%%%%%%%%%%%%%%%%%%%%%%%%%%%%%%%%%%%%%%%%%%%%%%%
%%%%%%%%%%%%%%%%%%%%%%%%%%%%%%%%%%%%%%%%%%%%%%%%%%%%%%%%%%%%%%%%%
%%%%%%%%%%%%%%%%%%%%%%%%%%%%%%%%%%%%%%%%%%%%%%%%%%%%%%%%%%%%%%%%%

\subsection{Parametric  EDE}\label{sec:pEDE}

%%%%%%%%%%%%%%%%%%%%%%%%%%%%%%%%%%%%%%%%%%%%%%%%%%%%%%%%%%%%%%%%%
%%%%%%%%%%%%%%%%%%%%%%%%%%%%%%%%%%%%%%%%%%%%%%%%%%%%%%%%%%%%%%%%%
%%%%%%%%%%%%%%%%%%%%%%%%%%%%%%%%%%%%%%%%%%%%%%%%%%%%%%%%%%%%%%%%%
%%%%%%%%%%%%%%%%%%%%%%%%%%%%%%%%%%%%%%%%%%%%%%%%%%%%%%%%%%%%%%%

\begin{figure*}
\centering
\includegraphics[scale=0.60]{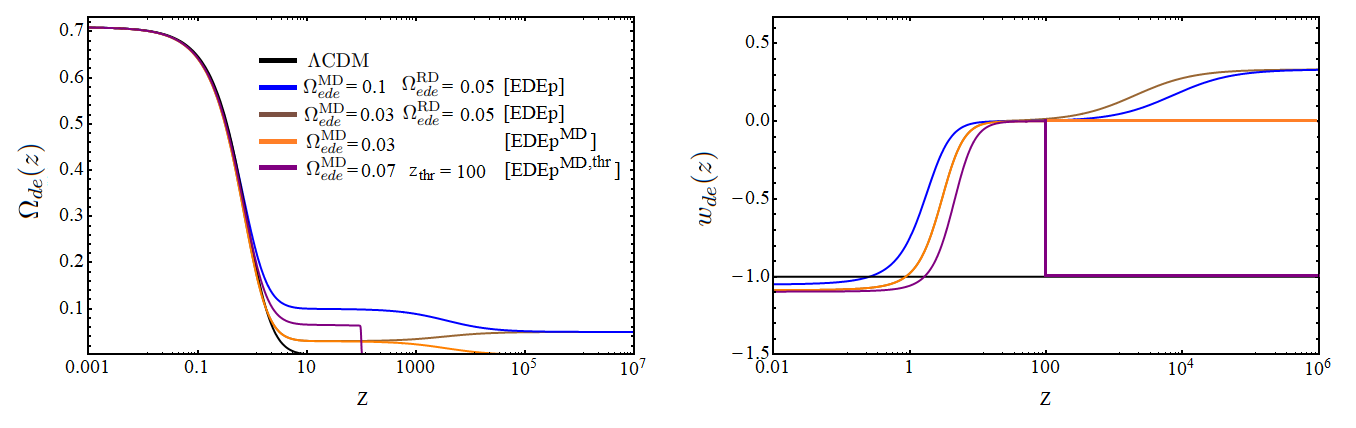}
\caption{Functions $\Omega_{de}(z)$ (left plot) and $w_{de}(z)=p_{de}(z)/\rho_{de}(z)$ (right plot) obtained using the parametrizations described in Sec. \ref{sec:pEDE}, for some illustrative values of the parameters.}
\label{fig:param}
\end{figure*}

%%%%%%%%%%%%%%%%%%%%%%%%%%%%%%%%%%%%%%%%%%%%%%%%%%%%%%%%%%%%%%%

We are interested in building a simple parameterization of the DE density that allows us to reproduce the behavior of uncoupled quintessence models with scaling or attractor solutions, while keeping a certain degree of generality and flexibility. The presence of scaling or attractor solutions is one of the most interesting features of DE models, as it allows to retrieve the current acceleration phase starting from a wide range of initial conditions. They represent a class of DE models that are promising to alleviate the fine tuning issues that affect a cosmological constant. For this reason, we want our parameterization to be able to generate two plateaux in $\Omega_{de}(z)=\rho_{de}(z)/\rho_{c}(z)$, with $\rho_{de}(z)$ and $\rho_{c}(z)=3H^2(z)/8\pi G$ the dark energy and critical energy densities in the Universe, respectively. The first plateau occurs in the radiation-dominated epoch (RDE), and the second one in the matter-dominated era (MDE). These plateaux can possibly have different heights. This is what happens for instance in quintessence models with a single exponential potential $V(\phi)=V_0e^{-\sqrt{8\pi G}\lambda\phi}$  \cite{Wetterich:1987fm,Copeland:1997et,Barreiro:1999zs}, where the fractions of EDE in the RDE and MDE depend only on one parameter, $\lambda$ (cf. Appendix A). 

With this aim in mind, we generalize the parameterizations proposed earlier in \cite{Pettorino:2013ia,Planck:2015bue} and we consider here a DE density with the following form,
\begin{equation}\label{eq:rhoDE}       
\rho_{de}(z)=\rho_{1} (1+z)^{4}+\rho_2 (1+z)^{3} +\rho_3 (1+z)^{3(1+w)}\,,
\end{equation}
parameterized by the constant energy densities $\rho_1$, $\rho_2$ and $\rho_3$  and by a constant equation of state parameter (EoS) $w$. We call this parameterization EDEp, where the "p" reminds us of the `plateaux' that characterise it. The last term of \eqref{eq:rhoDE} is able to mimic the behavior of a late-time dynamical DE with the $w$CDM form \cite{Turner:1997npq}, whereas the first two terms produce the  plateaux in the RDE and MDE. It is useful to write the constants $\rho_1$ and $\rho_2$ in terms of dimensionless parameters, as follows 
\begin{equation}\label{eq:dimensionalesschi}
\rho_1=\chi_1\Omega^{(0)}_{r,*}\rho^{(0)}_c\qquad ;\qquad \rho_2=\chi_2\Omega^{(0)}_{m,*}\rho^{(0)}_c\,,
\end{equation}
where $\Omega^{(0)}_{r,*}$ and $\Omega^{(0)}_{m,*}$ are the current density parameters of radiation and matter, respectively, computed considering three massless neutrinos\footnote{Notice that $\Omega^{(0)}_{r,*}$ and $\Omega^{(0)}_{m,*}$ are not equal to $\Omega^{(0)}_{r}$ and $\Omega^{(0)}_{m}$. The latter are the true radiation and matter density parameters, since they take into account the non-relativistic behavior of the massive neutrino of $0.06$ eV in the late-time universe that we have considered in all our numerical analyses. $\Omega^{(0)}_{r,*}$ and $\Omega^{(0)}_{m,*}$ have been defined for practical purposes, to allow an easier definition of the EDE fractions in the RDE and MDE in terms of the dimensionless parameters $\chi_1$ and $\chi_2$, respectively, cf. \eqref{eq:fractions}. This simplifies the implementation of our EDEp parameterization in the Einstein-Boltzmann code, see Sec. \ref{sec:Method}. The relative difference between $\Omega^{(0)}_{m}$ and $\Omega^{(0)}_{m,*}$ is $\sim 0.4\%$ for a total neutrino mass of $0.06$ eV, whereas the one between $\Omega^{(0)}_{r}$ and $\Omega^{(0)}_{r,*}$ is $\sim 13\%$.}, and $\rho_c^{(0)}\equiv\rho_c(z=0)$ is the current critical energy density. The superscripts $(0)$ denote quantities evaluated at present. In this work we consider standard General Relativity (GR) and a flat Friedmann-Lema\^itre-Robertson-Walker universe. If we want the DE density to be positive during  the cosmic expansion we have to demand $\chi_1,\chi_2,\rho_3>0$. Notice that deep in the RDE and MDE the fractions of energy in the Universe contained in the form of (early) DE are constant and given by
\begin{equation}\label{eq:fractions}
\Omega^{\rm RD}_{ede} = \frac{\chi_1}{1+\chi_1}\qquad ;\qquad \Omega^{\rm MD}_{ede} = \frac{\chi_2}{1+\chi_2}\,.
\end{equation}
The present dark energy density, $\rho^{(0)}_{de}$, can be directly computed from the Hubble parameter, $H_0=100h$ km/s/Mpc, and the reduced CDM and baryon density parameters, $\omega_{cdm}=\Omega_{cdm}^{(0)}h^2$ and $\omega_{b}=\Omega_{b}^{(0)}h^2$. Thus, one of the three $\rho_i$'s appearing in \eqref{eq:rhoDE} can be  expressed in terms of the other two, e.g. $\rho_3 = \rho^{(0)}_{de}-\rho_1-\rho_2$. It is clear then that in this EDE parametrization we deal with three additional parameters with respect to the $\Lambda$CDM, so we have nine cosmological parameters in total, namely the spectral index $n_s$ and amplitude $A_s$ of the primordial power spectrum, $H_0$, $\omega_b$, $\omega_{cdm}$, the optical depth to reionization, $\tau$, together with $w$, $\chi_1$ and $\chi_2$ (or, equivalently, $\Omega^{\rm RD}_{ede}$ and $\Omega^{\rm MD}_{ede}$). We consider the minimal configuration with a massive neutrino of $0.06$ eV and two massless neutrinos. 

For typical (low) values of $\Omega^{\rm RD}_{ede}$ and $\Omega^{\rm MD}_{ede}$ much smaller than one the DE fraction in EDEp can be approximated by

\begin{equation}\label{eq:OmegaDE}
\Omega_{de}(z)=\frac{\Omega^{\rm MD}_{ede}+\Omega^{\rm RD}_{ede}\left(\frac{1+z}{1+z_{eq}}\right)+\left(\frac{\Omega_{de}^{(0)}}{1-\Omega_{de}^{(0)}}-\Omega^{\rm MD}_{ede}\right)(1+z)^{3w}}{1+\frac{1+z}{1+z_{eq}}+\frac{\Omega_{de}^{(0)}}{1-\Omega_{de}^{(0)}}(1+z)^{3w}}\,,
\end{equation}
with $z_{eq}$ the redshift at the matter-radiation equality time.

As the DE fluid under consideration is covariantly conserved, i.e. it does not interact with the other species, we can make use of the following equation to compute its associated pressure,

\begin{align}\label{eq:pDE}
\dot{\rho}_{de}+&3H(\rho_{de}+p_{de})=0\\
& \longrightarrow p_{de}(z)=\frac{\rho_1}{3}(1+z)^{4}+w\rho_3(1+z)^{3(1+w)}\,,\nonumber
\end{align}
where the dot refers to a derivative with respect to the cosmic time. The corresponding equation of state parameter can be obtained from the ratio of \eqref{eq:pDE} and \eqref{eq:rhoDE}, i.e. $w_{de}(z)=p_{de}(z)/\rho_{de}(z)$. It is clear that $w_{de}=1/3$ and $w_{de}=0$ in the RDE and MDE, respectively, and $w_{de}\approx w$ at present for low values of $\chi_2$. For the perturbations, in our main analyses we take the sound speed of the DE fluid to be equal to the speed of light in the DE rest frame \cite{Ballesteros:2010ks}, i.e. $\hat{c}_s=1$, so in our model the DE does not cluster efficiently. We will study also what happens when we allow for values of $\hat{c}_s<1$. Our simple parametrization is able to reproduce the phenomenology of a quintessence model with a potential energy density that reduces to an exponential function of the scalar field in the RDE and MDE. These exponential potentials are controlled by two different constant parameters, which allow for the two independent plateaux of $\Omega_{ede}$. See Appendix \ref{sec:appendixA} for details. 

It is useful to compare our parametrization \eqref{eq:rhoDE} or just \eqref{eq:OmegaDE} with those previously studied in \cite{Doran:2006kp,Calabrese:2011hg,Pettorino:2013ia,Planck:2015bue}. If we set $\Omega^{\rm RD}_{ede}=\Omega^{\rm MD}_{ede}$, the phenomenology of \eqref{eq:rhoDE} is almost identical to the one found in the EDE1 parametrization of \cite{Pettorino:2013ia} (originally proposed by \cite{Doran:2006kp} and also employed by \cite{Calabrese:2011hg,Planck:2015bue}), and when $w$ is set to $-1$ also to the EDE2 parametrization of \cite{Pettorino:2013ia}. If we set $\Omega^{\rm RD}_{ede}=0$ in \eqref{eq:rhoDE} EDE is completely negligible during the RDE,  

\begin{equation}\label{eq:EDEp2}
\rho_{de}(z)=\rho_2 (1+z)^{3} +\rho_3 (1+z)^{3(1+w)}\,.
\end{equation}
We denote this particular case of EDEp as EDEp$^{\rm MD}$, to remind us that the plateau is in this case following the matter component. The EDE3 parametrization of \cite{Pettorino:2013ia,Planck:2015bue} can be retrieved to a good approximation if we set $\Omega^{\rm RD}_{ede}=0$ and $w=-1$ and turn the second term on in \eqref{eq:rhoDE} at a particular `threshold' redshift $z_{\rm thr}$, in the MDE. This is what we call EDEp$^{{\rm MD},thr}$: 

\begin{equation}\label{eq:EDEp2zthr}
\rho_{de}(z)=\rho_2 (1+z)^{3}\theta(z_{{\rm thr}}-z) +\rho_3 (1+z)^{3(1+w)}\,, 
\end{equation}
with $\theta$ the Heaviside step function. For $z_{\rm thr}\to\infty$ we recover the pure EDEp$^{\rm MD}$ parametrization \eqref{eq:EDEp2}. We show typical shapes of the functions $\Omega_{de}(z)$ and  $w_{de}(z)$ obtained with EDEp, EDEp$^{\rm MD}$ and EDEp$^{{\rm MD},thr}$ in Fig. 1. Finally, the behavior of the EDE4 parametrization of \cite{Pettorino:2013ia} can be also reproduced if we set $\Omega^{\rm MD}_{ede}=0$ and $w=-1$ and deactivate the first term in \eqref{eq:rhoDE} at a specific value of the scale factor in the RDE. Thus, with \eqref{eq:rhoDE} we can study the same scenarios as in \cite{Pettorino:2013ia,Planck:2015bue}, and also analyze what happens when we do not only have one single plateau, but two, one in the RDE and another one in the MDE. Hence, we will be able to compare the fitting results obtained with our parameterization with those reported in previous works in the literature, and also appreciate the evolution of the constraints on the EDE fraction with the data in the last years. In addition, by allowing departures of $w$ from $-1$ we can also see what is the effect of the late-time dynamics of DE on the fitting results. On top of this, we also explore different values of the EDE sound speed $\hat{c}_s$. This shows that our parameterization includes  previous ones but is more general.

Coupled quintessence models with an interaction between dark energy and dark matter can possess scaling solutions too, see  \cite{Amendola:1999er}, and \cite{Gomez-Valent:2020mqn} for recent constraints. In some running vacuum models (RVMs) with a vacuum energy density $\rho_{\rm vac}=\frac{3}{8\pi G}(c_0+\nu H^2)$ there is a plateau with $\Omega_{\rm vac}^{\rm RD}=\Omega_{\rm vac}^{\rm MD}\simeq\nu$ \cite{SolaPeracaula:2017esw}, and in the so-called Ricci RVMs, in which $\rho_{\rm vac}=\frac{3}{8\pi G}(c_0+\nu R/12)$ with $R$ the Ricci scalar, the vacuum energy density follows the one of non-relativistic matter during the RDE and MDE, with $\Omega_{\rm vac}^{\rm MD}\simeq\nu/4$ during the MDE \cite{SolaPeracaula:2021gxi}. In these models, though, the pressure of the dark energy/vacuum component is different from the one of the dominant species in the radiation- and matter-dominated eras, in contrast to what happens in the uncoupled quintessence models described before, and there are also important differences at the linear perturbations level. Differences in perturbations are also found in Brans-Dicke cosmology, but in this case both, the energy density and pressure of the effective DE fluid, evolve as for relativistic and non-relativistic matter in the RDE and MDE, respectively, for reasonable values of the Brans-Dicke parameter $\omega_{\rm BD}$ \cite{SolaPeracaula:2019zsl,SolaPeracaula:2020vpg}. Some of these models are able to loosen the cosmological tensions, see the aforementioned references for further details. Here we focus in the non-interacting scenarios with scaling behavior discussed above, assuming GR. We present the constraints on EDEp \eqref{eq:rhoDE}, EDEp$^{\rm MD}$ \eqref{eq:EDEp2} and EDEp$^{{\rm MD},thr}$ \eqref{eq:EDEp2zthr} in Sec. \ref{sec:resultsP}.

%%%%%%%%%%%%%%%%%%%%%%%%%%%%%%%%%%%%%%%%%%%%%%%%%%%%%%%%%%%%%%%%%
%%%%%%%%%%%%%%%%%%%%%%%%%%%%%%%%%%%%%%%%%%%%%%%%%%%%%%%%%%%%%%%%%
%%%%%%%%%%%%%%%%%%%%%%%%%%%%%%%%%%%%%%%%%%%%%%%%%%%%%%%%%%%%%%%%%

\subsection{Tomographic EDE}\label{sec:npEDE}

We further consider the possibility of binning the amount of $\Omega_{de}(z)$, similarly to what previously explored in \cite{Pettorino:2013ia,Planck:2015bue}, and also in our EDEp$^{\rm MD,thr}$ parametrization. While in such cases, though, only the fraction of dark energy below the threshold redshift at which it would become non negligible was let free to vary, here we constrain the EDE density in different bins; we then perform a tomographic analysis using the data sets described in Sec. \ref{sec:Data}. For convenience we fix 11 bins, as indicated in Table \ref{tableBins}: the right column identifies the amplitude within each of them. At low redshifts, $z < 5$, we assume a $w$CDM behavior \cite{Turner:1997npq}. Here, again, $\rho_{de}^{(0)}$ can be directly determined from $H_0$, $\omega_{b}$ and $\omega_{cdm}$; the other constant densities $\rho_{i}$, with $i=A,...,J$ (cf. Table \ref{tableBins}), are left free in our Monte Carlo runs, together with $w$ and the six usual $\Lambda$CDM parameters. We keep $\hat{c}_s=1$. Our main aim is to see how much EDE we can have in each bin, and therefore which shape of $\Omega_{de}(z)$ is preferred by the data, regardless of its complexity. Our binned method is similar (but not equal) to the one employed in \cite{Hojjati:2013oya}. The corresponding fitting results and reconstructed shapes of $\Omega_{de}(z)$ obtained with different data set combinations are shown and  discussed in Sec. \ref{sec:resultsNP}.

\begin{table*}[t!]
\centering
\begin{tabular}{|c  |c |}
\hline
{\small Redshift bin} & {\small $\rho_{de}(z)$}  
\\\hline
$z\leq 5$  & $\rho_{de}^{(0)}(1+z)^{3(1+w)}$
\\\hline
$5<z\leq 10$  & $\rho_{A}(1+z)^{3}$
\\\hline
$10<z\leq 50$  & $\rho_{B}(1+z)^{3}$
\\\hline
$50<z\leq 200$  & $\rho_{C}(1+z)^{3}$
\\\hline
$200<z\leq 500$  & $\rho_{D}(1+z)^{3}$
\\\hline
$500<z\leq 1000$  & $\rho_{E}(1+z)^{3}$
\\\hline
$1000<z\leq 2000$  & $\rho_{F}(1+z)^{4}$
\\\hline
$2000<z\leq 3000$  & $\rho_{G}(1+z)^{4}$
\\\hline
$3000<z\leq 5000$  & $\rho_{H}(1+z)^{4}$
\\\hline
$5000<z\leq 10^4$  & $\rho_{I}(1+z)^{4}$
\\\hline
$ z>10^{4}$  & $\rho_{J}(1+z)^{4}$
\\\hline
\end{tabular}
\caption{Binning of the EDE density employed in the tomographic analysis described in Sec \ref{sec:npEDE}.}
\label{tableBins}
\end{table*}

%%%%%%%%%%%%%%%%%%%%%%%%%%%%%%%%%%%%%%%%%%%%%%%%%%%%%%%%%%%%%%%%%
%%%%%%%%%%%%%%%%%%%%%%%%%%%%%%%%%%%%%%%%%%%%%%%%%%%%%%%%%%%%%%%%%
%%%%%%%%%%%%%%%%%%%%%%%%%%%%%%%%%%%%%%%%%%%%%%%%%%%%%%%%%%%%%%%%%

\section{Methodology}\label{sec:Method}

We have implemented the various parametrizations of Sec. \ref{sec:pEDE} and also the binned $\rho_{de}(z)$ described in Sec. \ref{sec:npEDE} in our own modified version of the Einstein-Boltzmann code \texttt{CLASS} \cite{Blas:2011rf}. We have used it to compute the background history and solve the linear perturbations equations, which is of course crucial to obtain all the theoretical quantities that we need to confront our models to the cosmological data, see Sec. \ref{sec:Data}. We have put constraints on the parameters of our models through the usual Bayesian exploration of the parameter space. Its high dimensionality is not only due to the pure cosmological parameters entering the models, but also to the nuisance parameters that are required to model the theory and experiment systematics, as it happens e.g. with the CMB foregrounds (cf. the references in Sec. \ref{sec:CMB}). Such large parameter spaces must be explored with Monte Carlo techniques as the Metropolis-Hastings algorithm \cite{Metropolis:1953aaa,Hastings:1970bbb}. For this purpose we have employed the Monte Carlo sampler \texttt{MontePython} \cite{Audren:2012wb}. We have used flat priors for the cosmological parameters in common with the $\Lambda$CDM model, with widths that fully respect the {\it Planck} 2018 uncertainties \cite{Planck:2018vyg}. Regarding the priors on the EDE fractions, previous studies in the literature showed that they are always lower than $\sim 10\%$ \cite{Pettorino:2013ia,Planck:2015bue}, so we have used the following flat prior on $\chi_i$ (with $i=1,2$) \eqref{eq:dimensionalesschi} in our Monte Carlo analyses of the parametrizations of Sec. \ref{sec:pEDE}: $0<\chi_i<0.12$; and, of course, similar priors have been imposed in each bin of the tomographic analysis of Sec. \ref{sec:npEDE}.

%%%%%%%%%%%%%%%%%%%%%%%%%%%%%%%%%%%%%%%%%%%%%%%%%%%%%%%%%%%%%%%%%
%%%%%%%%%%%%%%%%%%%%%%%%%%%%%%%%%%%%%%%%%%%%%%%%%%%%%%%%%%%%%%%%%
%%%%%%%%%%%%%%%%%%%%%%%%%%%%%%%%%%%%%%%%%%%%%%%%%%%%%%%%%%%%%%%%%

\section{Data}\label{sec:Data}

%%%%%%%%%%%%%%%%%%%%%%%%%%%%%%%%%%%%%%%%%%%%%%%%%%%%%%%%%%%%%%%%%

\subsection{Description of the individual data sets}\label{sec:individualDS}

In this section we list the various data sets used in this study. We explain the most important features of each of them and provide the corresponding references.

%%%%%%%%%%%%%%%%%%%%%%%%%%%%%%%%%%%%%%%%%%%%%%%%%%%%%%%%%%%%%%%%%
\subsubsection{Cosmic microwave background}\label{sec:CMB}

We consider the full {\it Planck} 2018 TTTEEE+lowE likelihood \cite{Planck:2018vyg}. It includes the data on the CMB temperature (TT) and polarization (EE) anisotropies, and also their cross-correlations (TE) at low and high multipoles. We call this data set, in short, CMBpol. In some of our fits we also include the data on CMB lensing, i.e.  we consider the  {\it Planck} 2018 TTTEEE+lowE+lensing likelihood \cite{Planck:2018vyg,Planck:2018lbu}. We denote the resulting CMB data set as CMBpolens.

%%%%%%%%%%%%%%%%%%%%%%%%%%%%%%%%%%%%%%%%%%%%%%%%%%%%%%%%%%%%%%%%%

\subsubsection{Supernovae of Type Ia}\label{sec:SNIa}

We use the observed apparent magnitude and redshifts from the standardized 1048 SNIa of the Pantheon compilation \cite{Scolnic:2017caz}. We have duly taken into account the existing correlations among them through the corresponding covariance matrix. The absolute magnitude of these SNIa, $M$, is left free in the fitting analysis, and we impose a prior on it, see Sec. \ref{sec:SH0ES} for details. 

%%%%%%%%%%%%%%%%%%%%%%%%%%%%%%%%%%%%%%%%%%%%%%%%%%%%%%%%%%%%%%%%%

\subsubsection{Prior on $M$}\label{sec:SH0ES}

There exist a $\sim 4.1\sigma$ tension between the values of the Hubble parameter measured by SH0ES \cite{Riess:2020fzl}, 

\begin{equation}\label{eq:priorH0}
H_{0,\rm SH0ES}=(73.2\pm 1.3)\,{\rm km/s/Mpc}\,,
\end{equation} 
and the one inferred from the CMB anisotropy maps by {\it Planck} 2018 under the assumption of the flat $\Lambda$CDM model \cite{Planck:2018vyg}, $H_0=(67.36\pm 0.54)$ km/s/Mpc, \cite{Verde:2019ivm,DiValentino:2020zio}. It is still unclear whether this tension is due to some sort of new physics beyond the standard model \cite{DiValentino:2021izs} or to the presence of systematic errors in the data, see e.g. \cite{SPT:2017sjt,Efstathiou:2020wxn,Mortsell:2021nzg,Freedman:2021ahq}, but in this work we want to study its status in the context of the EDE models described in Sec. \ref{sec:EDE}, assuming that the tension has a physical origin. We do so by using in some of our fitting analyses the SH0ES effective calibration prior on the absolute magnitude of the SNIa as provided in \cite{Camarena:2019moy},

\begin{equation}\label{eq:priorM}
M_{\rm SH0ES}=-19.2191\pm 0.0405\,.
\end{equation}
The latter is obtained from the calibration of nearby SNIa (at $z\lesssim 0.01$) with Cepheids \cite{Riess:2019cxk}. As explained in \cite{Camarena:2019moy}, it is better to use this prior rather than the one on $H_0$ \cite{Reid:2019tiq,Riess:2020fzl}, especially when it is combined with data from supernovae compilations that include the same SNIa in the Hubble flow considered by the SH0ES team. This is the case e.g. of the Pantheon compilation, which we employ in all our fits, see Sec. \ref{sec:SNIa} and \ref{sec:combinedDS}. Hence, using the prior on $M$ \eqref{eq:priorM} instead of $H_0$ \eqref{eq:priorH0} we avoid unwanted double-counting issues. In any case, the results obtained from the two priors is almost indistinguishable for models with no abrupt features at very low redshift ($z\lesssim 0.1$) \cite{Benevento:2020fev,Camarena:2021jlr}, as the ones we analyze in this work or those of \cite{Gomez-Valent:2020mqn,SolaPeracaula:2020vpg}. In some of our Tables we also provide the best-fit values of $M$. This allows us to quantity the `$M$ tension', i.e. the tension between the latter and \eqref{eq:priorM}. We show in Sec. \ref{sec:resultsNP} that the statistical level of the SH0ES-{\it Planck} tension can be in some cases quite different when formulated in terms of the Hubble parameter and the absolute magnitude of SNIa.

%%%%%%%%%%%%%%%%%%%%%%%%%%%%%%%%%%%%%%%%%%%%%%%%%%%%%%%%%%%%%%%%%
        
\subsubsection{Baryon acoustic oscillations}\label{sec:BAO}

The interactions between photons and baryons before the decoupling of the former at $z_{dec}\sim 1100$ caused the so-called baryon acoustic oscillations (BAO). Their imprint can be observed as a peak in the two-point correlation function of matter fluctuations at a scale of about $147$ Mpc, which can be used as a standard ruler to constrain cosmological models if one assumes that it is not partially caused e.g. by any peculiar feature in the primordial power spectrum \cite{2dFGRS:2005yhx,SDSS:2005xqv}. Several galaxy surveys have been able to provide precise data on BAO, either in terms of the dilation scale $D_V$,
\begin{equation}
\frac{D_V(z)}{r_d}=\frac{1}{r_d}\left[D_M^2(z)\frac{cz}{H(z)}\right]^{1/3}\,,
\end{equation}
with $D_M=(1+z)D_{A}(z)$ being the comoving angular diameter distance and $r_d$ the sound horizon at the baryon drag epoch, or even by splitting (when possible) the transverse and line-of-sight BAO information and hence being able to provide data on $D_{A}(z)/r_d$ and $H(z)r_d$ separately, with some degree of correlation. The surveys provide the values of the measurements at some effective redshift(s). We employ the following BAO data points in this work:

\begin{itemize}

\item $D_V/r_d$ at $z=0.122$ provided in \cite{Carter:2018vce}, which combines the dilation scales previously reported by the 6dF Galaxy Survey (6dFGS) \cite{Beutler:2011hx} at $z=0.106$ and the one obtained from the Sloan Digital Sky Survey (SDSS) Main Galaxy Sample at $z=0.15$ \cite{Ross:2014qpa}.

\item The anisotropic BAO data measured by BOSS using the LOWZ ($z=0.32$) and CMASS ($z=0.57$) galaxy samples \cite{Gil-Marin:2016wya}.

\item The dilation scale measurements by WiggleZ at $z=0.44,0.60,0.73$ \cite{Kazin:2014qga}. The galaxies contained in the WiggleZ catalog are located in a patch of the sky that partially overlaps with those present in the CMASS sample by BOSS. Nevertheless, the two surveys are independent, work under different seeing conditions, instrumental noise, etc. and target different types of galaxies. The correlation between the CMASS and WiggleZ data has been quantified in \cite{Beutler:2015tla}, where the authors estimated the correlation coefficient to be $\lesssim 2\%$. This justifies the inclusion of the WiggleZ data in our analysis, although their statistical weight is much lower than those from BOSS and in practice their use does not have any important impact on our results. 

\item $D_A/r_d$ at $z=0.81$ measured by the Dark Energy Survey (DES) \cite{DES:2017rfo}.

\item  The anisotropic BAO data from the extended BOSS Data Release 16 (DR16) quasar sample at $z=1.48$ \cite{Neveux:2020voa}.

\item The combined measurement of the anisotropic BAO information obtained from the Ly$\alpha$ absorption and quasars of the final data release (SDSS DR16) of eBOSS, at $z=2.334$ \cite{duMasdesBourboux:2020pck}.

\end{itemize}

%%%%%%%%%%%%%%%%%%%%%%%%%%%%%%%%%%%%%%%%%%%%%%%%%%%%%%%%%%%%%%%%%

\subsubsection{Weak lensing}\label{sec:WL}

In some of our fitting analyses we employ a prior on the composite quantity $S_8=\sigma_8(\Omega_m^{(0)}/0.3)^{0.5}$, with $\sigma_8$ the root-mean-square ({\it rms}) of mass fluctuations at scales of $8h^{-1}$ Mpc. More concretely, we use the prior

\begin{equation}\label{eq:priorS8}
S_8 = 0.762^{+0.025}_{-0.024}\,,
\end{equation}
obtained from the combined tomographic weak gravitational lensing (WL) analysis of the Kilo Degree Survey (KiDS +VIKING-450) and the Dark Energy Survey (DES-Y1) \cite{Joudaki:2019pmv}. This value is in tension with the {\it Planck} 2018 TT,TE,EE+lowE +lensing $\Lambda$CDM best-fit value ($S_8=0.832\pm 0.013$) at the level of $\sim 2.5\sigma$. Other weak lensing studies find similar results, see e.g. \cite{Joudaki:2017zdt,Wright:2020ppw,Heymans:2020gsg}. They point to a lower amount of LSS formation in the Universe than the one that can be accommodated by the concordance model. This tension is usually formulated directly in terms of the $\sigma_8$ parameter when information from galaxy clustering measurements is considered \cite{Macaulay:2013swa,SolaPeracaula:2017esw,Nesseris:2017vor,Heymans:2020gsg}, e.g. using data from redshift-space distortions (RSD), see Sec. \ref{sec:RSD}. The prior \eqref{eq:priorS8} has been obtained under the strong model assumption of the flat $\Lambda$CDM, so strictly speaking it can only be used in fitting analyses of the concordance model. Nevertheless, we use it in some of our fitting runs too in order to study the ability of our EDE models to loosen the $S_8$/$\sigma_8$ tension, and the impact of the WL data on our constraints on the EDE fraction.

\begin{table*}[t!]
\centering
\begin{tabular}{|c  ||c | c | c | c |   }
\hline
\multicolumn{5}{|c|}{CMBpol+SNIa}
\\\hline
{\small Parameter} & {\small $\Lambda$CDM}  & {\small $w$CDM} & {\small EDEp}  &  {\small EDEp$^{\rm MD}$}
\\\hline
{\small $\omega_b$}  & {{\small$0.02239^{+0.00014}_{-0.00015}$}}  & {{\small$0.02237^{+0.00016}_{-0.00015}$}} &  {{\small$0.02238^{+0.00017}_{-0.00016}$}} &  {{\small$0.02234^{+0.00015}_{-0.00016}$}}
\\\hline
{\small $\omega_{cdm}$} & {{\small$0.1199^{+0.0014}_{-0.0013}$}}  & {{\small$0.1204\pm 0.0014$}}  & {{\small$0.1218^{+0.0016}_{-0.0015}$}}  & {{\small$0.1208^{+0.0014}_{-0.0015}$}}
\\\hline
{\small $\tau$} & {{\small$0.055^{+0.007}_{-0.008}$}} & {{\small$0.054\pm 0.008$}} & {{\small$0.054\pm 0.008$}}  &  {{\small$0.053^{+0.007}_{-0.008}$}} 
\\\hline
{\small $n_s$} & {{\small$0.9659\pm 0.0044$}}  & {{\small$0.9646^{+0.0044}_{-0.0045}$}}  & {{\small$0.9642^{+0.0044}_{-0.0045}$}}  & {{\small$0.9642^{+0.0043}_{-0.0046}$}}  
\\\hline
{\small $H_0$ [km/s/Mpc]} & {{\small$67.60^{+0.59}_{-0.61}$}} & {{\small$68.55^{+1.12}_{-1.10}$}} & {{\small$68.71\pm 1.16$}} & {{\small$68.61^{+1.09}_{-1.16}$}}  
\\\hline
{\small$\sigma_8$}  & {{\small$0.811^{+0.007}_{-0.008}$}}  & {{\small$0.823\pm 0.014$}} & {{\small$0.817^{+0.014}_{-0.015}$}}  &   {{\small$0.818^{+0.015}_{-0.014}$}}
\\\hline
$r_d$ [Mpc] & {\small$147.02^{+0.28}_{-0.32}$} & {\small$146.92\pm 0.30$} & {\small$146.18^{+0.66}_{-0.43}$} & {\small$146.75^{+0.36}_{-0.30}$}
\\\hline
{\small $w$} & {\small$-1$}  & {{\small$-1.039^{+0.035}_{-0.039}$}}  & {{\small$-1.050^{+0.041}_{-0.040}$}}   & {{\small$-1.053^{+0.038}_{-0.042}$}}  
\\\hline
$\Omega^{\rm RD}_{ede}\,(\%)$ & {\small$0$}  & {\small$0$}  & {\small$<0.91\,(<2.08)$ }    & {\small$0$}
\\\hline
$\Omega^{\rm MD}_{ede}\,(\%)$ & {\small$0$}  & {\small$0$}  & {\small$<0.27\,(<0.69)$} &  {\small$<0.29\,(<0.69)$}
\\\hline
\multicolumn{5}{|c|}{CMBpolens+SNIa+$M$}
\\\hline
{\small Parameter} & {\small $\Lambda$CDM}  & {\small $w$CDM} & {\small EDEp}  & {\small EDEp$^{\rm MD}$}
\\\hline
{\small $\omega_b$}  & {{\small$0.02257^{+0.00015}_{-0.00014}$}}  & {{\small$0.02241^{+0.00014}_{-0.00016}$}} &  {{\small$0.02245^{+0.00015}_{-0.00019}$}} &{{\small$0.02239\pm 0.00015 $}}
\\\hline
{\small $\omega_{cdm}$} & {{\small$0.1179\pm 0.0012$}}  & {{\small$0.1200\pm 0.0015$}}  & {{\small$0.1212\pm 0.0016$}}  &
{{\small$0.1206\pm 0.0012$ }}
\\\hline
{\small $\tau$} & {{\small$0.057^{+0.007}_{-0.008}$}} & {{\small$0.054\pm 0.008$}} &  {{\small$0.055\pm 0.008$}} &
{{\small$0.056^{+0.007}_{-0.008} $}}                   
\\\hline
{\small $n_s$} & {{\small$0.9709\pm 0.0044$}}  & {{\small$0.9658^{+0.0044}_{-0.0047}$}}  & {{\small$0.9659^{+0.0045}_{-0.0044}$}}  &
{{\small$0.9656^{+0.0041}_{-0.0042} $}}
\\\hline
{\small $H_0$ [km/s/Mpc]} & {{\small$68.56^{+0.56}_{-0.54}$}} & {{\small$70.55^{+0.86}_{-0.88}$}} &   {{\small$70.63^{+0.86}_{-0.82}$}} &
{{\small$70.20^{+0.52}_{-0.69} $}}
\\\hline
{\small$\sigma_8$}  & {{\small$0.811^{+0.007}_{-0.008}$}}  & {{\small$0.838\pm 0.014$}} &  {{\small$0.830^{+0.014}_{-0.013}$}} &
{{\small$0.835\pm 0.010$}}
\\\hline
$r_d$ [Mpc] & {\small$147.36^{+0.28}_{-0.29}$} & {\small$146.98^{+0.33}_{-0.30}$} & {\small$146.21^{+0.73}_{-0.44}$}  &
{{\small $146.79\pm 0.29$}}
\\\hline
{\small $w$} & {\small$-1$}  & {{\small$-1.098^{+0.035}_{-0.032}$}}  & 
{{\small$-1.099^{+0.034}_{-0.032}$}} &
{{\small$-1.099^{+0.030}_{-0.025}$}}
\\\hline
$\Omega^{\rm RD}_{ede}\,(\%)$ & {\small$0$}  & {\small$0$}  & {\small$<1.14\,(2.44)$}   & 0
\\\hline
$\Omega^{\rm MD}_{ede}\,(\%)$ & {\small$0$}  & {\small$0$}  & {\small$<0.22\,(0.52)$} &
{\small$<0.22\,(0.54)$}
\\\hline
\end{tabular}
\caption{The mean fit values and $68.3\%$ confidence limits for the $\Lambda$CDM, the $w$CDM, EDEp \eqref{eq:rhoDE} and EDEp$^{\rm MD}$ \eqref{eq:EDEp2}, using the CMBpol+SNIa and CMBpolens+SNIa+$M$ data sets (cf. Sec. \ref{sec:Data}). For $\Omega^{\rm RD}_{ede}$ and $\Omega^{\rm MD}_{ede}$ we also show the $2\sigma$ limits inside the parentheses. See the comments in Sec. \ref{sec:resultsP}.}
\label{table2}
\end{table*}

Some works in the literature \cite{Poulin:2018cxd,Smith:2019ihp,Murgia:2020ryi,Smith:2020rxx} have reported significant hints in favor of an ultra-light axion-like EDE model in which the Hubble function is larger than in the $\Lambda$CDM during the pre-recombination epoch due to the presence of a scalar field in the Universe with an associated constant and very large potential energy density that at $z\sim 3000-5000$ decays as fast as (or even faster than) radiation. This excess of energy, which can be of about $\sim 7-10\%$ at the maximum of $\Omega_{ede}$, produces a decrease of the sound horizon $r_d$ that has to be compensated by an increase of $H_0$ to keep the location of the first peak of the CMB temperature power spectrum intact. These models lead, though, to much larger values of $\omega_{cdm}$, and this generates also a larger growth of matter perturbations in the late-time universe. Hence, the $H_0$ tension in this model is alleviated, but at the expense of worsening the $S_8/\sigma_8$ one \cite{Hill:2020osr,Ivanov:2020ril,DAmico:2020ods}. Something similar happens in the New EDE model proposed in \cite{Niedermann:2019olb,Niedermann:2020dwg} or the early modified gravity model studied in \cite{Braglia:2020auw}. Although the $S_8$/$\sigma_8$ tension is not as statistically significant as the $H_0$ one (cf. Sec. \ref{sec:SH0ES}), it is interesting to check whether it is possible for our EDE models to alleviate both tensions at the same time. This could happen, in principle, since our models have some features that are not present in \cite{Poulin:2018cxd,Niedermann:2020dwg,Niedermann:2019olb,Smith:2019ihp,Hill:2020osr,Ivanov:2020ril,Murgia:2020ryi,Smith:2020rxx,DAmico:2020ods}. In particular, our models also allow for: (i) a non-negligible fraction of EDE during the MDE that could help to slow down the formation of structures in the Universe due to the inability of EDE to cluster; and (ii) a late-time dynamical dark energy component, which can also help in this direction.

\cite{Sanchez:2020vvb} has raised some concerns about the use of $\sigma_8$ and derived quantities as $S_8$. He suggests the use of $\sigma_{12}$, defined as the {\it rms} linear theory variance at the fixed scale of $12$ Mpc, and $S_{12}=\sigma_{12}(\omega_{m}/0.14)^{0.4}$. We provide the values of these parameters in some of our Tables, together with the usual $\sigma_8$ and $S_8$. We will show in Sec. \ref{sec:resultsNP} that the enhancement in the LSS formation processes observed in previous works \cite{Hill:2020osr,Ivanov:2020ril,DAmico:2020ods} for different EDE models is also found for those described in Sec. \ref{sec:EDE} when $\sigma_{12}$ and $S_{12}$ are considered, instead of the usual $\sigma_8$ and $S_8$. We will see, though, that this issue can be mitigated by allowing the presence of EDE in the MDE, after the decoupling of the CMB photons.

%%%%%%%%%%%%%%%%%%%%%%%%%%%%%%%%%%%%%%%%%%%

\subsubsection{Redshift-space distortions and direct peculiar velocity measurements}\label{sec:RSD}

LSS Information can be also extracted from statistical measurements of the anisotropic clustering of galaxies in redshift space, and from the direct measurement of peculiar velocities when redshift-independent distance indicators are available. Galaxy redshift surveys obtain constraints on the product of the growth rate of structure, $f(z)=\frac{d\ln \delta_m(a)}{d\ln a}$, and $\sigma_8(z)$. These are the measurements that we include in our data set:

\begin{itemize}

\item  The data point at $z=0.035$ obtained from the joint analysis of 6dFGS and SDSS peculiar velocities \cite{Said:2020epb}.

\item The two data points provided by the Galaxy and Mass Assembly survey (GAMA) at $z=0.18$ \cite{Simpson:2015yfa} and $z=0.38$ \cite{Blake:2013nif}.

\item The four points at $z=0.22,0.41,0.60,0.78$ measured by WiggleZ \cite{Blake:2011rj}.

\item  The RSD measurements by BOSS from the power spectrum and bispectrum of the DR12 galaxies contained in the LOWZ ($z=0.32$) and CMASS ($0.57$) samples \cite{Gil-Marin:2016wya}.

\item The two points at $z=0.60,0.86$ reported by the VIMOS Public Extragalactic Redshift Survey (VIPERS) \cite{Mohammad:2018mdy}.

\item The point at $z=0.77$ by VIMOS VLT Deep Survey (VVDS) \cite{Guzzo:2008ac,Song:2008qt}. 

\item The Subaru FMOS galaxy redshift survey (FastSound) measurement at $z=1.36$ \cite{Okumura:2015lvp}.

\item The measurement by eBOSS DR16 at $z=1.48$ \cite{Neveux:2020voa}.

\end{itemize}

We call this data set RSD in short because most of these points are obtained from the analysis of redshift-space distortions. Whenever this data set is combined with the BAO data we take into account the correlations reported in \cite{Gil-Marin:2016wya,Neveux:2020voa}.

%%%%%%%%%%%%%%%%%%%%%%%%%%%%%%%%%%%%%%%%%%%%%%%%%%%%%%%%%%%%%%%%%

\subsection{Combined data sets}\label{sec:combinedDS}

We list here the data set combinations that we have used in this work. We have built these combinations from the individual data sets of Sec. \ref{sec:individualDS}. The corresponding fitting results are presented and  discussed in Sec. \ref{sec:results}.

In view of the significance of the $H_0$ tension, we deem it is of utmost importance to study and compare the constraints on our EDE models obtained: (i) with a minimal data set composed by CMBpol+SNIa; and (ii) adding on top of the latter the SH0ES prior on the absolute magnitude of SNIa, i.e. using CMBpol+SNIa+$M$. The SNIa data help to break the strong degeneracies found in the $w$-$H_0$ plane when only CMB data are used in the analysis (see e.g. \cite{Alestas:2020mvb}), and this is why we add the Pantheon compilation to the {\it Planck} 2018 data in these minimal data configurations. Their comparison allows us to quantify the impact of the SH0ES prior on the fitting results and on the ability of the models to loosen the $H_0$/$M$ tension.

The properties and limitations of the EDEp and EDEp$^{\rm MD}$ parametrizations are already grasped with the aforementioned minimal data sets (cf. Table \ref{table2}). For EDEp$^{{\rm MD},thr}$ we study also the effect of the CMB lensing and BAO+RSD data when combined with CMBpol+SNIa. We provide the corresponding constraints in figures 2 and 3. For the analyses of the binned $\rho_{de}(z)$ described in Sec. \ref{sec:npEDE} we report our results in Table \ref{table3}, where we explicitly test the impact of BAO and the weak lensing data, by considering not only the minimal data sets described in the previous paragraph, but also adding the information on BAO and BAO+WL. In addition, we redo the fitting analyses considering also the CMB lensing in order to quantify its impact. Tables \ref{table4} and \ref{table5} complement the discussion of our tomographic study.

%%%%%%%%%%%%%%%%%%%%%%%%%%%%%%%%%%%%%%%%%%%%%%%%%%%%%%%%%%%%%%%

\section{Results and discussion}\label{sec:results}

We present and discuss now the results obtained from the fitting analyses of the EDE parametrizations and the tomographic EDE described in Secs. \ref{sec:pEDE} and \ref{sec:npEDE}, respectively, using the data sets listed in the previous section.

\subsection{Results for the parametric analysis}\label{sec:resultsP}

The mean fit values and corresponding uncertainties for the various  cosmological parameters in the EDEp and EDEp$^{\rm MD}$ parametrizations obtained with the baseline CMBpol+SNIa dataset are reported in Table \ref{table2}. The constraints on the fraction of early dark energy in the EDEp parametrization in the radiation- and matter-dominated epochs, $\Omega^{\rm RD}_{ede}$ and $\Omega^{\rm MD}_{ede}$, are very strong. They lie below $\sim 2\%$ and $\sim 0.7\%$, respectively, at the $2\sigma$ c.l. It is interesting to observe that the upper limit of $\Omega^{\rm MD}_{ede}$ in the EDEp$^{\rm MD}$ parametrization coincides with the one obtained in the more general EDEp. The constraints on $\Omega^{\rm MD}_{ede}$ and $\Omega^{\rm RD}_{ede}$ in EDEp are quite independent. Actually, we have checked that the correlation coefficient between these two parameters is pretty small, $\sim 5.6\%$. As already noticed in \cite{Pettorino:2013ia}, the low upper limits on $\Omega^{\rm RD}_{ede}$ and $\Omega^{\rm MD}_{ede}$ are due to the very tight constraint on the fraction of EDE around the CMB decoupling time. The latter acts as an anchor for $\Omega^{\rm RD}_{ede}$ and even more for $\Omega^{\rm MD}_{ede}$, since in the last scattering surface the matter energy density is already $\sim 3$ times larger than the radiation one.

Another result from Table \ref{table2} to remark is that EDEp cannot alleviate significantly the $H_0$ and $\sigma_8$ tensions. The shape of the early dark energy density seems to be too restricted in these parametrization. There is a slight increase of $H_0$ in EDEp and EDEp$^{\rm MD}$ with respect to the $\Lambda$CDM, but it is mainly due to the dynamics of the late-time DE, and this is why the major part of the effect is already found with the $w$CDM parametrization \cite{Turner:1997npq}.

%%%%%%%%%%%%%%%%%%%%%%%%%%%%%%%%%%%%%%%%%%%%%%%%%%%%%%%%%%%%%%%%%
%%%%%%%%%%%%%%%%%%%%%%%%%%%%%%%%%%%%%%%%%%%%%%%%%%%%%%%%%%%%%%%%%

\begin{figure}[t!]
\begin{center}
\label{fig:cs2}
\includegraphics[width=3.2in, height=2.5in]{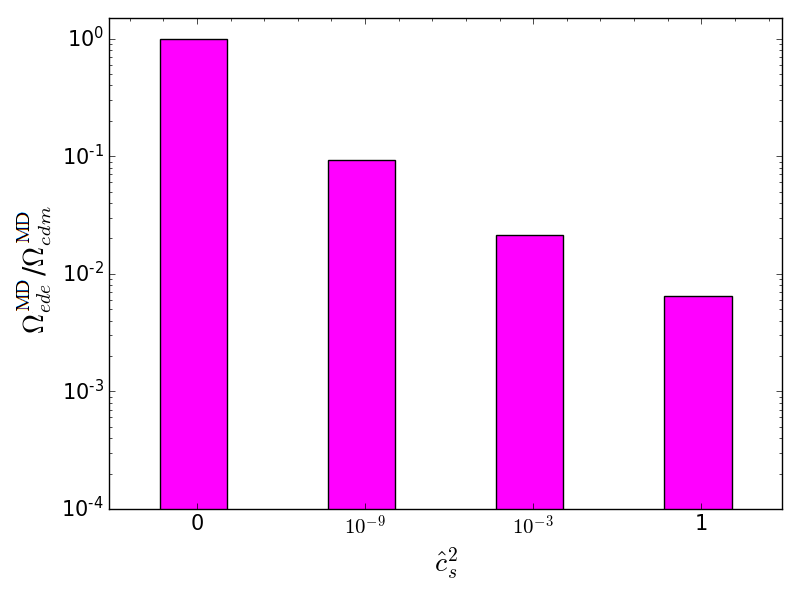}
\caption{Constraints at $2\sigma$ c.l. on the fraction of EDE in the MDE as a function of its sound speed $\hat{c}_s$ and under the CMBpol+SNIa data set. We show these constraints through the ratio $\Omega^{\rm MD}_{ede}/\Omega^{\rm MD}_{cdm}$, with $\Omega^{\rm MD}_{cdm}$ the fraction of CDM during the MD, given by $\Omega^{MD}_{cdm}=\omega_{cdm}/[(1+\chi_2)(\omega_{cdm}+\omega_b)]$ up to a very small correction due to massive neutrinos. We have employed here the EDEp$^{\rm MD}$ parametrization \eqref{eq:EDEp2} with $w=-1$. When $\hat{c}_s\to 0$ we find that the constraints on $\Omega^{\rm MD}_{ede}$ loosen and approach to those obtained for cold dark matter, as expected, since in this limit the DE behaves exactly as CDM during the MDE and there exists an obvious degeneracy between these two parameters.}
\end{center}
\end{figure}

%%%%%%%%%%%%%%%%%%%%%%%%%%%%%%%%%%%%%%%%%%%%%%%%%%%%%%%%%%%%%%%

When we include the SH0ES prior in our fitting analysis we increase, of course, the value of the Hubble parameter, see Table \ref{table2}. The tension with the distance ladder determination is now only of $\sim 1.7\sigma$, but this is again mainly thanks to the lowering of $w$\footnote{The upper bound of $\Omega^{\rm RD}_{ede}$ in the EDEp parametrization increases by $0.36\%$ with respect to when we do not include the SH0ES prior, whereas the value of $\Omega^{\rm MD}_{ede}$ decreases by  $0.17\%$ (cf. Table \ref{table2}).  The increase of $\Omega^{\rm RD}_{ede}$ helps to alleviate the $H_0$ tension, although this effect is only marginal. It forces the decrease of $\Omega^{\rm MD}_{ede}$ to fulfill the strong constraint on the fraction of EDE around the CMB decoupling time.}, which now lies more in the phantom region ($3\sigma$ away below $w=-1$) \cite{Banerjee:2020xcn}. The values of $w$ and $H_0$ are almost identical to those found in the $w$CDM parametrization, and EDE does not have any important impact on the $H_0$ tension in the context of the EDEp parametrization. The loosening of the $H_0$ tension is accompanied by a slight worsening of the $\sigma_8$ one due to the positive correlation between $H_0$ and $\sigma_8$. Phantom dark energy leads to lower values of the DE density in the past and this produces, in turn, an increase of the structure formation processes in the Universe. It seems that the EDE density has a too restricted form in the parameterizations under study here. They allow for plateaux and generalize previous studies, but still seem to be quite constrained and unable to resolve the tensions. In the next section we then investigate the binned tomographic approach described in Sec. \ref{sec:npEDE}. Before moving on, however, we discuss briefly two options to weaken the constraints on the fraction of EDE.

%%%%%%%%%%%%%%%%%%%%%%%%%%%%%%%%%%%%%%%%%%%%%%%%%%%%%%%%%%%%%%%
\begin{figure}[t!]
\begin{center}
\label{fig:FMD}
\includegraphics[width=3.5in, height=2.5in]{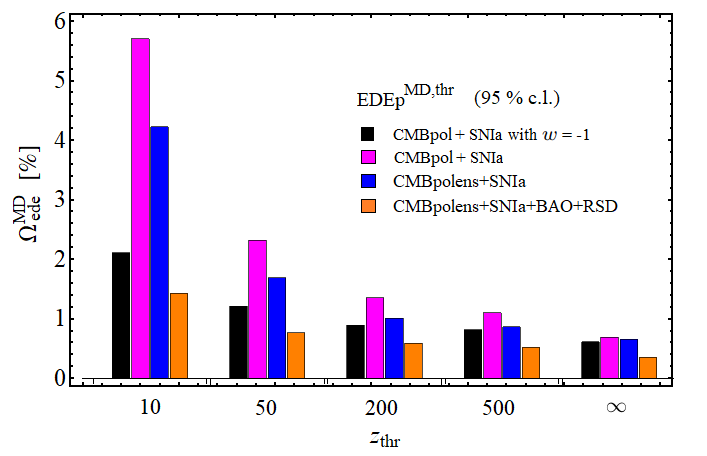}
\caption{Constraints on $\Omega^{\rm MD}_{ede}$ at $95\%$ c.l. obtained from the fitting analyses of the EDEp$^{\rm MD,thr}$ parametrization \eqref{eq:EDEp2zthr}, and under the data sets shown in the legend.}
\end{center}
\end{figure}
%%%%%%%%%%%%%%%%%%%%%%%%%%%%%%%%%%%%%%%%%%%%%%%%%%%%%%%%%%%%%%%

%%%%%%%%%%%%%%%%%%%%%%%%%%%%%%%%%%%%%%%%%%%%%%%%%%%%%%%%%%%%%%%
\begin{figure*}[t!]
\begin{center}
\label{fig:SomeConstraintsEDEp2}
\includegraphics[width=6in, height=4in]{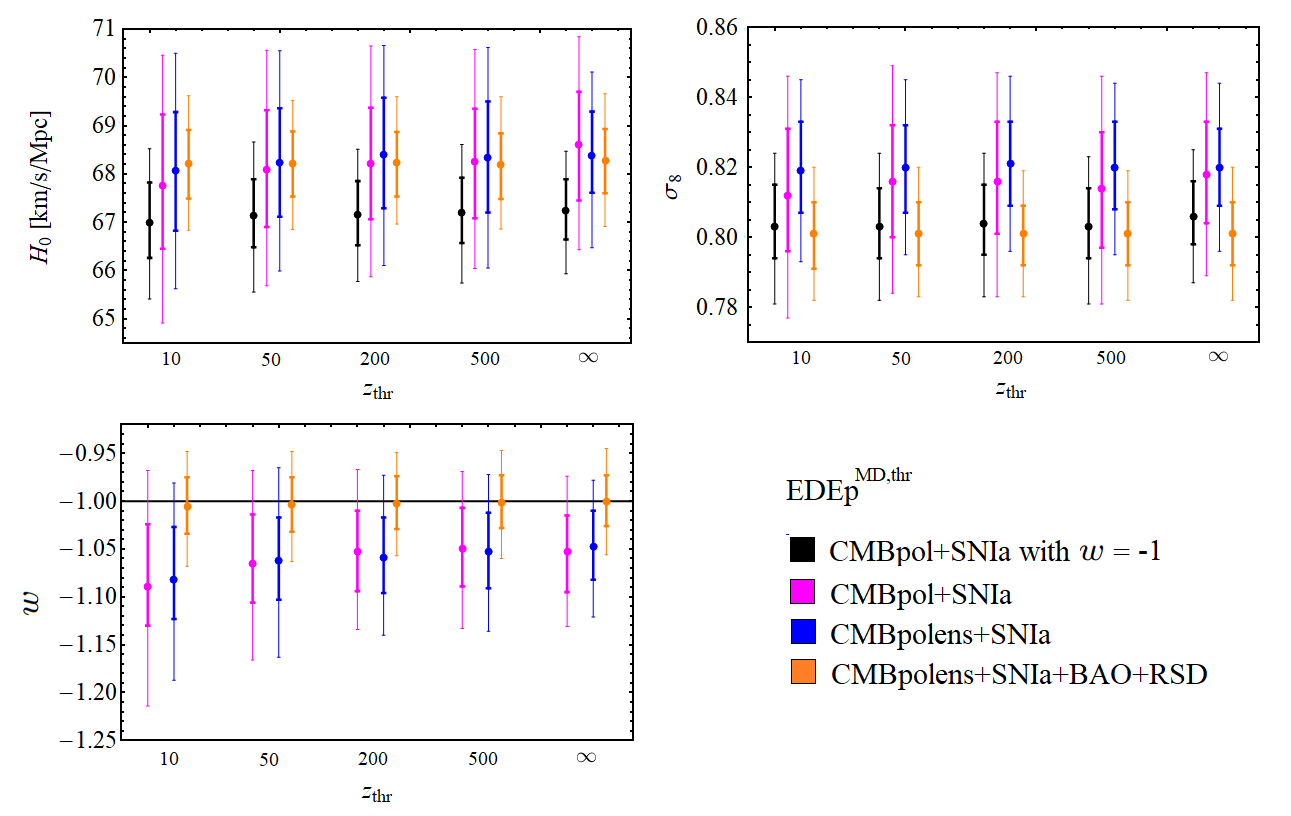}
\caption{Constraints for $H_0$, $\sigma_8$ and $w$ at $1\sigma$ and $2\sigma$ c.l. obtained with the EDEp$^{{\rm MD},thr}$ parametrization \eqref{eq:EDEp2zthr} for different values of the threshold redshift $z_{thr}$ and using the same combined data sets as in Fig. 3.}

\end{center}

\end{figure*}
%%%%%%%%%%%%%%%%%%%%%%%%%%%%%%%%%%%%%%%%%%%%%%%%%%%%%%%%%%%%%%%

Much weaker constraints on the fraction of EDE can be obtained by allowing for values of $\hat{c}^2_s\ll 1$. The only difference between our EDE and cold dark matter during the MDE is found at the perturbations level. Both are pressureless fluids at the background level, but have different $\hat{c}_s$. The latter is equal to 1 for EDE and 0 for the dark matter. Hence, we expect the constraints on $\Omega^{\rm MD}_{ede}$ to loosen if we decrease the value of $\hat{c}_s$ of EDE, and even obtain a full degeneracy between the fraction of EDE and CDM during the MDE in the limit $\hat{c}_s\to 0$. This is actually what happens, as we explicitly show in Fig. 2. We have to say, though, that the change in the sound speed does not help to alleviate the cosmological tensions, neither. For instance, under the CMBPol+SNIa data set the Hubble parameter remains close to $\sim 67.5$ km/s/Mpc, and the $1\sigma$ uncertainty is lower than $\sim 0.8$ km/s/Mpc regardless of the value of $\hat{c}_s$ under consideration when the late-time DE dynamics is switched off.

One can also get weaker constraints on EDE in the MDE by activating EDE at lower redshifts. In order to study this effect we explore the EDEp$^{{\rm MD},thr}$ parametrization \eqref{eq:EDEp2zthr}. In Fig. 3 we provide the $2\sigma$ c.l. bounds on $\Omega^{\rm MD}_{ede}$ obtained with the baseline data set CMBpol+SNIa with and without late-time dark energy dynamics, and also adding the CMB lensing and the BAO+RSD data sets. When $z_{\rm thr}\to \infty$ we recover the constraints obtained in the EDEp$^{\rm MD}$ model, of course, but when we allow for lower values of the threshold redshift (below the CMB decoupling one) we get larger upper bounds on $\Omega^{\rm MD}_{ede}$, which depend on the concrete data set under consideration and also on $z_{\rm thr}$. We report the results obtained with $z_{\rm thr}=10,50,200,500,\infty$. The addition of the {\it Planck} 2018 CMB lensing to the CMBpol+SNIa baseline data set leads to stronger constraints on $\Omega^{\rm MD}_{ede}$. Its value decreases by a $\sim 25\%$ $\forall{z_{\rm thr}}$. When we also include the BAO+RSD data the decrease is even bigger, $\sim 75\%$. If we turn off the late-time dynamics of DE we also obtain tighter bounds on $\Omega^{\rm MD}_{ede}$, just because in this case we remove the degeneracy between this parameter and $w$. Our Fig. 3 can be compared with Figs. 6-7 of \cite{Pettorino:2013ia} and Fig. 11 of \cite{Planck:2015bue}, which were obviously obtained with older data sets. In \cite{Pettorino:2013ia} the authors employed CMB data from WMAP9 combined with small scale measurements from the South Pole Telescope (SPT), whereas in \cite{Planck:2015bue} the authors employed the {\it Planck} 2015 CMB likelihood and studied the impact of some other background and weak lensing data sets, as described in Sec. 4 in that reference. The results presented in this section constitute a significant update obtained with the {\it Planck} 2018 likelihood and also other more recent background and LSS data (cf. Sec. \ref{sec:Data} for details).

\begin{table*}[t!]
\centering
\resizebox{\textwidth}{!}{\begin{tabular}{|c  ||c | c | c | c | c |}
\hline
{\small Parameter} & {\small CMBpol+SNIa}  & {\small CMBpol+SNIa+$M$} &  {\small CMBpol+SNIa+$M$+BAO} & {\small CMBpol+SNIa+$M$+BAO+WL}
\\\hline
{\small $\omega_b$}  & {{\small$0.02257^{+0.00022}_{-0.00023}$}}  & {{\small$0.02277^{+0.00023}_{-0.00025}$}} &  {{\small$0.02282^{+0.00024}_{-0.00025}$}}           &
{{\small$0.02259\pm 0.00021$}}
\\\hline
{\small $\omega_{cdm}$} & {{\small$0.1222^{+0.0020}_{-0.0021}$}}  & {{\small$0.1221\pm 0.0022$}}  & {{\small$0.1223^{+0.0020}_{-0.0021}$}}  &
{{\small$0.1200^{+0.0013}_{-0.0014}$}}
\\\hline
{\small $\tau$} & {{\small$0.055\pm 0.009$}} & {{\small$0.056\pm 0.009$}} &  {{\small$0.057^{+0.008}_{-0.009}$}}                  &
{{\small$0.053\pm 0.008$}}
\\\hline
{\small $n_s$} & {{\small$0.9727^{+0.0072}_{-0.0073}$}}  & {{\small$0.9752^{+0.0067}_{-0.0069}$}}  & {{\small$0.9760^{+0.0070}_{-0.0071}$}}  &
{{\small$0.9740^{+0.0061}_{-0.0067}$}}
\\\hline
{\small $H_0$ [km/s/Mpc]} & {{\small$68.29^{+1.26}_{-1.36}$}} & {{\small$70.86^{+1.00}_{-1.10}$}} &   {{\small$70.38^{+0.84}_{-0.89}$}}         &
{{\small$69.85^{+0.76}_{-0.77}$}}
\\\hline
{\small $M$} & {{\small$-19.405\pm0.032$}}  &
{{\small$-19.342^{+0.023}_{-0.024}$}}                &       
{{\small$-19.350^{+0.019}_{-0.020}$}}                            &
{{\small$-19.365^{+0.018}_{-0.017}$}}
\\\hline
{\small$\sigma_8$}  & {{\small$0.854^{+0.023}_{-0.025}$}}  & {{\small$0.880^{+0.025}_{-0.029}$}} &  {{\small$0.877^{+0.026}_{-0.029}$}}      &
{{\small$0.833^{+0.016}_{-0.017}$}}
\\\hline
{\small$S_8$}  & {{\small $0.869^{+0.025}_{-0.029}$}} & {{\small $0.863\pm 0.029$}} &  {{\small $0.866^{+0.026}_{-0.028}$}}                                     &
{{\small $0.819^{+0.014}_{-0.016}$}}
\\\hline
{\small$\sigma_{12}$}  & {{\small$0.840^{+0.019}_{-0.022}$}}  & {{\small$0.843\pm 0.022$}} &  {{\small$0.843^{+0.020}_{-0.022}$}}               &
 {{\small$0.806^{+0.012}_{-0.013}$}}
\\\hline
{\small$S_{12}$}  & {{\small $0.851^{+0.022}_{-0.025}$}} & {{\small $0.854^{+0.024}_{-0.027}$}} &  {{\small $0.855^{+0.023}_{-0.026}$}}      &
 {{\small$0.810^{+0.014}_{-0.016}$}}
\\\hline
$r_d$ [Mpc]& {\small$145.66^{+0.90}_{-0.70}$} & {{\small$145.22^{+0.95}_{-0.90}$}} & {{\small$145.06^{+0.97}_{-0.89}$}}       &
{{\small$146.51^{+0.64}_{-0.51}$}}
\\\hline                          
{\small $w$} & {\small$-1.037^{+0.043}_{-0.041}$}  & {{\small$-1.070\pm 0.038$}}  &  {{\small$-1.048^{+0.037}_{-0.034}$}}       &
{{\small$-1.037^{+0.032}_{-0.031}$}}

\\\hline\hline

{\small Parameter} & {\small CMBpolens+SNIa}  & {\small CMBpolens+SNIa+$M$} & {\small CMBpolens+SNIa+$M$+BAO} & {\small CMBpolens+SNIa+$M$+BAO+WL}
\\\hline
{\small $\omega_b$}  & {{\small$0.02257^{+0.00021}_{-0.00023}$}}  & {{\small$0.02274^{+0.00023}_{-0.00025}$}} &  {{\small$0.02277^{+0.00022}_{-0.00024}$}} & {{\small$0.02266^{+0.00021}_{-0.00020}$}}
\\\hline
{\small $\omega_{cdm}$} & {{\small$0.1215\pm0.0016$}}  & {{\small$0.1211\pm 0.0017$}}  & {{\small$0.1214\pm 0.0016$}}  & {{\small$0.1193^{+0.0013}_{-0.0014}$}}
\\\hline
{\small $\tau$} & {{\small$0.054^{+0.008}_{-0.009}$}} & {{\small$0.056^{+0.008}_{-0.009}$}} &  {{\small$0.056\pm 0.008$}} & {{\small$0.055^{+0.008}_{-0.007}$}}
\\\hline
{\small $n_s$} & {{\small$0.9721^{+0.0063}_{-0.0071}$}}  & {{\small$0.9748^{+0.0067}_{-0.0071}$}}  & {{\small$0.9746^{+0.0068}_{-0.0069}$}}  & {{\small$0.9727^{+0.0061}_{-0.0068}$}}
\\\hline
{\small $H_0$ [km/s/Mpc]} & {{\small$68.25^{+1.27}_{-1.26}$}} & {{\small$70.84^{+1.04}_{-1.07}$}} &   {{\small$70.21^{+0.80}_{-0.84}$}} & {{\small$70.00^{+0.76}_{-0.73}$}}
\\\hline
{\small $M$} & {{\small$-19.407^{+0.031}_{-0.030}$}}                & 
{{\small$-19.343^{+0.024}_{-0.025} $}}                  & 
{{\small$-19.355\pm 0.018 $}}                  & 
{{\small $-19.362\pm 0.016$}}

\\\hline
{\small$\sigma_8$}  & {{\small$0.845^{+0.017}_{-0.020}$}}  & {{\small$0.868^{+0.019}_{-0.021}$}} &  {{\small$0.864^{+0.019}_{-0.021}$}}& {{\small$0.839^{+0.013}_{-0.015}$}}
\\\hline
{\small$S_8$}  & {{\small $0.858^{+0.019}_{-0.021}$}} & {{\small $0.848^{+0.020}_{-0.022}$}} &  {{\small $0.853^{+0.019}_{-0.020}$}} & {{\small $0.824^{+0.012}_{-0.013}$}}
\\\hline
{\small$\sigma_{12}$}  & {{\small$0.831^{+0.014}_{-0.015}$}}  & {{\small$0.831^{+0.015}_{-0.016}$}} &  {{\small$0.833^{+0.015}_{-0.016}$}}& {{\small$0.810^{+0.010}_{-0.011}$}}
\\\hline
{\small$S_{12}$}  & {{\small $0.841^{+0.016}_{-0.017}$}} & {{\small $0.840^{+0.017}_{-0.020}$}} &  {{\small $0.843\pm 0.018$}}   & {{\small $0.815^{+0.011}_{-0.013}$}} 
\\\hline
$r_d$ [Mpc] & {\small$145.94^{+0.69}_{-0.57}$} & {\small$145.65^{+0.82}_{-0.71}$} & {\small$145.51^{+0.83}_{-0.71}$} & {\small$146.46^{+0.56}_{-0.44}$}
\\\hline
{\small $w$} & {\small$-1.035^{+0.040}_{-0.039}$}  & {{\small$-1.067^{+0.037}_{-0.036}$}}  & {{\small$-1.050^{+0.034}_{-0.035}$}} & {{\small$-1.045\pm 0.032$}}
\\\hline
\end{tabular}}
\caption{The mean fit values and $68.3\%$ confidence limits for the most important cosmological (main+derived) parameters, obtained under different data sets (with and without CMB lensing) for the binned $\rho_{de}(z)$ described in Sec. \ref{sec:npEDE}. See Fig. 5 for the constraints on $\Omega_{de}(z)$, and Sec. \ref{sec:resultsNP} for a thorough discussion on these results.}
\label{table3}
\end{table*}

In Fig. 4 we provide the corresponding constraints on $H_0$, $\sigma_8$ and $w$ at $1\sigma$ and $2\sigma$ c.l. in the EDEp$^{\rm MD, thr}$ parameterization for the same scenarios explored in Fig. 3. They support some of the comments made in the previous paragraphs of this section, e.g. (i) the values of $\sigma_8$ and $H_0$ remain close to those found in the $\Lambda$CDM model. In other words, the tensions are not significantly alleviated in this class of scaling early dark energy models; (ii) phantom values of $w<-1$ allow us to decrease the $H_0$ tension very slightly; and (iii) larger values of $w$ lead to lower values of $\sigma_8$ due to the presence of a larger fraction of dark energy at low redshifts. This is why we get $w\sim -1$ when we include the BAO+RSD data set.

The dedicated analysis presented in this section  updated and generalized previous constraints on this class of early dark energy models, and  motivated the study of the next section, in which we will reconstruct the shape of $\Omega_{de}(z)$ without sticking to a restricted family of parametrizations. 

\subsection{Results for tomographic Dark Energy}\label{sec:resultsNP}

Now we provide the results obtained in the tomographic model described in Sec. \ref{sec:npEDE} in order to see whether more general shapes of $\Omega_{de}(z)$ can loosen the cosmological tensions. This is in fact suggested by previous analyses in the literature, see e.g. \cite{Poulin:2018cxd,Agrawal:2019lmo,Smith:2019ihp,Chudaykin:2020acu,Niedermann:2020dwg,Gogoi:2020qif,Niedermann:2019olb}. Our results are presented in Tables \ref{table3}-\ref{table5}, and also in Fig. 5. 

Table 3 and Fig. 5 confirm that a significant (and non-constant) fraction of EDE in the RDE can alleviate the $H_0$ tension if it can be kept below $\sim 0.6\%$ at $2\sigma$ c.l. around the CMB decoupling time, i.e. at $z\sim 1000-2000$. For instance, from the analysis of the CMBpolens+SNIa and CMBpolens+SNIa+$M$+BAO data sets we obtain $H_0=(68.25^{+1.27}_{-1.26})$ km/s/Mpc and $H_0=(70.21^{+0.80}_{-0.84})$ km/s/Mpc, respectively. They are $2.77\sigma$ and $1.95\sigma$ below the SH0ES measurement \eqref{eq:priorH0}, and the central value is significantly lower when the SH0ES prior is not considered. \cite{Hill:2020osr} reported similar results in the context of the ultra-light axion model. Nevertheless, if we compare the values of $M$ obtained from these data sets with the SH0ES value \eqref{eq:priorM} we still obtain a tension of $3.80\sigma$ and $3.07\sigma$, respectively. This means that in terms of $M$, the tension is bigger than when it is formulated in terms of $H_0$, and the capability of EDE of alleviating the tension is much lower, at least in the EDE framework we are considering here.

\begin{table*}[t!]
\centering
\resizebox{\textwidth}{!}{\begin{tabular}{|c  ||c | c | c | c | c |}
\hline
{\small Parameter} & {\small CMBpol+SNIa}  & {\small CMBpol+SNIa+$M$} &  {\small CMBpol+SNIa+$M$+BAO} & {\small CMBpol+SNIa+$M$+BAO+WL}
\\\hline
{\small $H_0$ [km/s/Mpc]} & {{\small$2.66\sigma$}} & {{\small$1.40\sigma$}} &   {{\small$1.81\sigma$}}         &
{{\small$2.22\sigma$}}
\\\hline
{\small $M$} & {{\small$3.60\sigma$}}  &
{{\small$2.62\sigma$}}                &       
{{\small$2.91\sigma$}}                            &
{{\small$3.31\sigma$}}
\\\hline
{\small$\sigma_8$}  & {{\small$2.89\sigma$}}  & {{\small$3.45\sigma$}} &  {{\small$3.32\sigma$}}      &
{{\small$2.65\sigma$}}
\\\hline
{\small$S_8$}  & {{\small $2.93\sigma$}} & {{\small $2.66\sigma$}} &  {{\small $2.85\sigma$}}                                     &
{{\small $1.98\sigma$}}
\\\hline
{\small$\sigma_{12}$}  & {{\small$2.66\sigma$}}  & {{\small$2.67\sigma$}} &  {{\small$2.73\sigma$}}               &
 {{\small$1.82\sigma$}}
\\\hline\hline

{\small Parameter} & {\small CMBpolens+SNIa}  & {\small CMBpolens+SNIa+$M$} & {\small CMBpolens+SNIa+$M$+BAO} & {\small CMBpolens+SNIa+$M$+BAO+WL}
\\\hline
{\small $H_0$ [km/s/Mpc]} & {{\small$2.73\sigma$}} & {{\small$1.41\sigma$}} &   {{\small$1.95\sigma$}} & {{\small$2.13\sigma$}}
\\\hline
{\small $M$} & {{\small$3.71\sigma $}}                & 
{{\small$2.62\sigma$}}                  & 
{{\small$3.07\sigma$}}                  & 
{{\small $3.28\sigma$}}
\\\hline
{\small$\sigma_8$}  & {{\small$2.96\sigma$}}  & {{\small$3.63\sigma$}} &  {{\small$3.50\sigma$}}& {{\small$3.03\sigma$}}
\\\hline
{\small$S_8$}  & {{\small $3.04\sigma$}} & {{\small $2.67\sigma$}} &  {{\small $2.91\sigma$}} & {{\small $2.25\sigma$}}
\\\hline
{\small$\sigma_{12}$}  & {{\small$2.69\sigma$}}  & {{\small$2.64\sigma$}} &  {{\small$2.71\sigma$}}& {{\small$2.05\sigma$}}

\\\hline
\end{tabular}}
\caption{Statistical significance of the cosmological tensions for the relevant parameters $H_0$, $M$, $\sigma_8$, $S_8$ and $\sigma_{12}$, obtained under different data sets (with and without CMB lensing) for the tomographic DE described in Sec. \ref {sec:npEDE}. The tensions are quantified by using the fitting values reported in Table \ref{table3} and the following measurements: the Hubble parameter \eqref{eq:priorH0} and the absolute magnitude of SNIa \eqref{eq:priorM} by SH0ES; the value of $\sigma_8$ = 0.760 $\pm$ 0.022 provided in \cite{Heymans:2020gsg}, which is obtained from the cosmic shear analysis by KiDS-1000 and the galaxy clustering data from BOSS. We use this value also to quantify the tension in $\sigma_{12}$, since $\sigma_{12}\simeq\sigma_{8}$ for the $\Lambda$CDM due to the low values of $H_0$ encountered in this model \cite{Sanchez:2020vvb}; and the WL measurement of $S_8$ \eqref{eq:priorS8}.}
\label{table4}
\end{table*}

We also see that the large fraction of $\Omega_{de}(z)$ required in the RDE to loosen the $H_0$ tension, which can be of about $\sim 4-5\%$ at $2\sigma$ c.l. according to some data sets that include the SH0ES prior on $M$ but no LSS information, leads to higher values of $\omega_{cdm}$ and, to a lesser extent, also of $n_s$, which in turn exacerbates the $\sigma_8$/$S_8$ tension. The former is needed to reduce the early integrated Sachs-Wolfe effect introduced by the EDE. This is aligned with previous works that also consider EDE in the pre-recombination epoch, see e.g. \cite{Hill:2020osr,Ivanov:2020ril,DAmico:2020ods,Murgia:2020ryi,Smith:2020rxx}. Tables \ref{table3} and \ref{table4} show that the tension decreases $\sim 1\sigma$ when it is analyzed through the LSS estimators $\sigma_{12}$ and $S_{12}$ \cite{Sanchez:2020vvb}. Nevertheless, it does not disappear (see also Table \ref{table5}). Under the CMBpolens+SNIa+$M$ data set the $\sigma_8$ tension is of $3.63\sigma$, whereas for the $\sigma_{12}$ parameter it is of $2.64\sigma$, and a similar decrease is observed when other data combinations are employed in the fitting analysis, see again Table \ref{table4}.

When the SH0ES prior on $M$ \eqref{eq:priorM} is taken into account in absence of LSS data, we get a $\sim 2\sigma$ evidence for the presence of a non-null EDE fraction during the RDE, as in \cite{Poulin:2018cxd}\footnote{There are some differences, though, between the ultra-light axion-like field model of \cite{Poulin:2018dzj,Poulin:2018cxd} and our reconstructed EDE. For instance, in the former the EDE sound speed is not constant and equal to one, but has a more complicated dependence on the scale factor and the wavemode $k$ \cite{Poulin:2018dzj}. Moreover, in the model of \cite{Poulin:2018dzj,Poulin:2018cxd} $\Omega_{de}(z)$ has a sizeable peak at $z\sim 3000-5000$, but the EDE fraction is negligible at much higher redshifts, whereas in our binned analysis the possibility of having a non-negligible fraction of EDE deep in the RDE is not excluded by the data. This is an interesting feature of EDE that obviously cannot be grasped with the model studied in \cite{Poulin:2018cxd,Poulin:2018dzj}, just because in that case $\rho_{de}(z)\approx const.$ by construction during radiation-domination, when the energy density of the scalar field is basically given by a flat (and very large) potential and EDE behaves as a cosmological constant. Thus, for large enough redshifts the EDE density in that model is always overcome by the one of radiation.}. Nevertheless, the inclusion of the weak lensing data from KiDS+VIKING-450 and  DES-Y1 (cf. Sec. \ref{sec:WL}) forces the EDE fraction in the RDE to be again compatible at $1\sigma$ with $0$ in order to allow $\omega_{cdm}$ to take values closer to the $\Lambda$CDM ones and not to worsen the $\sigma_8/S_8$ tension. Notice that, as expected, the upper bound on the fraction of EDE in the MDE that we obtain in our binned analysis is larger than the one found with EDEp and its variants, even when the prior on $S_8$ is not included (see Table \ref{table2} and Fig. 5). Indeed, higher fractions of EDE in the MDE (see again Fig. 5) also allow to keep the amount of LSS more under control. With the CMBpolens+SNIa+$M$+BAO+WL we obtain $H_0=(70.00^{+0.76}_{-0.73})$ km/s/Mpc and $M=-19.362\pm 0.016$. They are in $2.13\sigma$ and $3.28\sigma$ tension with the SH0ES values, respectively. Again, the tension in $M$ is larger than in $H_0$ by $\sim 1\sigma$. Regarding the LSS estimators, we obtain $S_8=0.824^{+0.012}_{-0.013}$ and $\sigma_{12}=0.810^{+0.010}_{-0.011}$. The former is in  $2.25\sigma$ tension with \eqref{eq:priorS8}, whereas the latter is compatible at $1\sigma$ with the value obtained in the $\Lambda$CDM. The tensions in $H_0$ and $S_8$ can be kept at $\sim 2\sigma$ c.l. under this concrete data set, as advocated by \cite{Murgia:2020ryi}.

%%%%%%%%%%%%%%%%%%%%%%%%%%%%%%%%%%%%%%%%%%%%%%%%%%%%%%%%%%%%%%%
\begin{figure*}[t!]
\begin{center}
\label{fig:OmegaEDE}
\includegraphics[width=6.7in, height=6.75in]{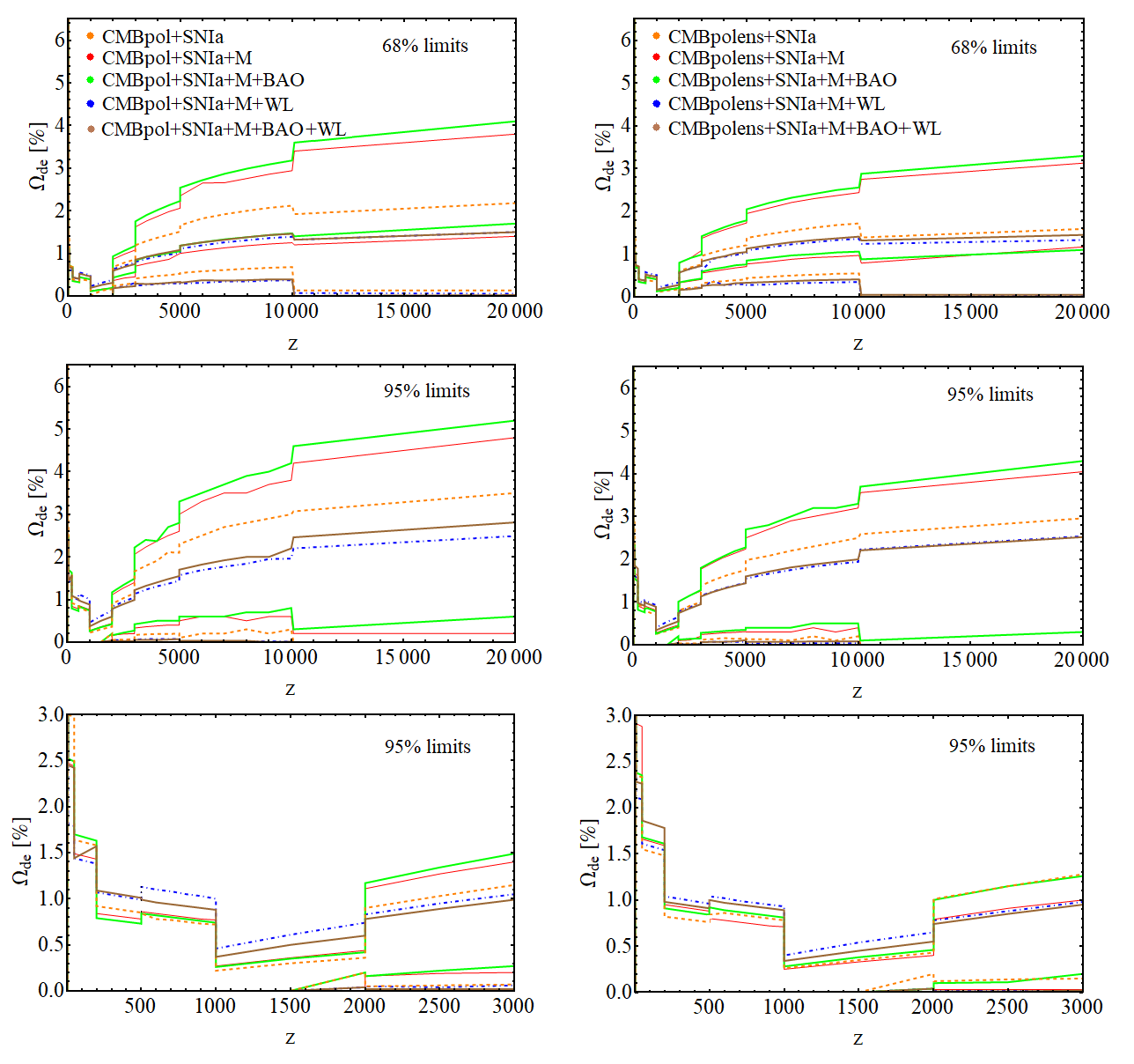}
\caption{{\it Left plots:} Reconstructed shapes of $\Omega_{de}(z)$ obtained from the fitting analyses without the Planck 2018 CMB lensing data (cf. Sec. \ref{sec:CMB}). In the first and second rows we show the constraints in the region $z\in [0,2\cdot 10^4]$ at $1\sigma$ and $2\sigma$ c.l., respectively. In the last row we zoom in the redshift range $z\in [0,3000]$ to better appreciate the details in the MDE. The tightest upper bound on $\Omega_{de}(z)$ is obtained around the CMB decoupling time, i.e. at $z\sim 1000-2000$, where the data force $\Omega_{de}(z)\lesssim 0.6\% $ at $2\sigma$ c.l.; {\it Right plots:} The same, but including the CMB lensing data. See the comments in Sec. \ref{sec:resultsNP}.}
\end{center}
\end{figure*}
%%%%%%%%%%%%%%%%%%%%%%%%%%%%%%%%
%%%%%%%%%%%%%%%%%%%%%%%%%%%%%%%

The effect of the CMB lensing from {\it Planck} 2018 on the EDE fraction can be appreciated by direct comparison of the plots in the left and right columns of Fig. 5. When the CMB lensing is included, the upper bound on $\Omega_{de}(z)$ is reduced by $\sim 1\%$ for $z\gtrsim 5000$,  by a $\sim 0.5\%$ for $3000\lesssim z\lesssim 5000$ and by a smaller fraction at lower redshifts. Some differences are found, though, depending on the other data sets employed in the fitting analyses. The preferred matter densities decrease, although they are still compatible at $1\sigma$ with the ones inferred without including the CMB lensing. This leads also to a slight decrease on the LSS estimators when the weak lensing prior is not considered. When the latter is included, the values of $\sigma_{12}$ and $S_{12}$ (and also $\sigma_{8}$ and $S_{8}$) remain stable under the addition of the CMB lensing likelihood.

We would like to remark the very low upper bounds that we obtain for the EDE fraction in the redshift range $z\in(1000,2000)$, cf. Fig. 5. The exact value for these upper bounds depend, of course, on the specific data set, e.g. the constraints are a little bit weaker when the WL prior \eqref{eq:priorS8} is used in the fitting analysis, but the EDE fraction when $\hat{c}_s=1$ is, in any case, very strongly constrained in that epoch and lies always below $0.6-0.7\%$ regardless of the data set under consideration. Under the full data set, the EDE fractions are constrained to be  below $2.6\%$ in the radiation-dominated epoch and $\lesssim 1-1.5\%$ in the redshift range $z\in (100,1000)$ at $2\sigma$ c.l. This limits strongly the possible impact of EDE on the value of the Hubble parameter or the present
amount of structure.

%%%%%%%%%%%%%%%%%%%%%%%%%%%%%%%%%%%%%%%%%%%%%%%%%%%%%%%%%%%%%%%%%
%%%%%%%%%%%%%%%%%%%%%%%%%%%%%%%%%%%%%%%%%%%%%%%%%%%%%%%%%%%%%%%%%
%%%%%%%%%%%%%%%%%%%%%%%%%%%%%%%%%%%%%%%%%%%%%%%%%%%%%%%%%%%%%%%%%

\section{Conclusions}\label{sec:conclusions}

In this work we have studied the phenomenological performance of a family of flexible parametrizations for the dark energy density that are able to mimic the scaling behavior that is encountered in a wide variety of quintessence models. DE has been treated as a perfect fluid with two plateaus in $\Omega_{de}(z)$, one in the matter-dominated epoch and another one in the radiation-dominated era. We have put very tight constraints on the fraction of EDE in the radiation- and matter-dominated epochs in the context of these models. The CMBpol+SNIa data set already forces these two quantities to lie below $2.44\%$ and $0.52\%$ at $2\sigma$ c.l., respectively. These strong constraints are necessary to respect the upper bound on the amount of EDE at the last scattering surface, as we have explicitly checked in our tomographic analysis. We have found that this class of scaling EDE models does not lead to a significant alleviation of the $H_0$ and $\sigma_8$ tensions. Larger EDE fractions are allowed by the data if: (i) the sound speed for DE is fixed at values $\hat{c}_s^2\ll 1$, since in this limit dark energy behaves as dark matter during the matter-dominated epoch at the background and perturbations levels and, hence, there is a huge degeneracy between these two components. If we would assume a vanishing sound speed for dark energy there would be no way to distinguish it from dark matter. We find indeed no bounds on the dark energy fraction in this case. This changes drastically already for a rather small sound speed of EDE of the order $10^{-4}$; and (ii) if EDE is switched on at later times, already in the matter-dominated era. Nevertheless, the cosmological tensions remain also in these cases. 

\begin{table*}[t!]
\centering
\begin{tabular}{|c  ||c | c | c | c  |}
\hline
{\small Parameter} & {\small $\Lambda$CDM}  & {\small $w$CDM} &  {\small Binned EDE} 
\\\hline
{\small $\omega_b$}  & {{\small$0.02255\pm 0.00013$}}  & {{\small$0.02246^{+0.00014}_{-0.00013}$}} &  {{\small$0.02282^{+0.00024}_{-0.00025}$}}         
\\\hline
{\small $\omega_{cdm}$} & {{\small$0.1183^{+0.0009}_{-0.0008}$}}  & {{\small$0.1197\pm 0.0010$}}  & {{\small$0.1223^{+0.0020}_{-0.0021}$}} 
\\\hline
{\small $\tau$} & {{\small$0.057^{+0.008}_{-0.009}$}} & {{\small$0.055^{+0.007}_{-0.008}$}} &  {{\small$0.057^{+0.008}_{-0.009}$}}             
\\\hline
{\small $n_s$} & {{\small$0.9699^{+0.0034}_{-0.0038}$}}  & {{\small$0.9665^{+0.0038}_{-0.0041}$}}  & {{\small$0.9760^{+0.0070}_{-0.0071}$}}  
\\\hline
{\small $H_0$ [km/s/Mpc]} & {{\small$68.49^{+0.35}_{-0.40}$}} & {{\small$68.78^{+0.66}_{-0.78}$}} &   {{\small$70.38^{+0.84}_{-0.89}$}}         
\\\hline
{\small $M$} & {{\small$-19.394^{+0.009}_{-0.011}$}}   &
{{\small$-19.371^{+0.013}_{-0.017}$}}                &       
{{\small$-19.350^{+0.019}_{-0.020}$}}                        
\\\hline
{\small$\sigma_8$}  & {{\small$0.807\pm{0.008}$}}  & {{\small$0.828^{+0.012}_{-0.013}$}} &  {{\small$0.877^{+0.026}_{-0.029}$}}     
\\\hline
{\small$S_8$}  & {{\small $0.807^{+0.012}_{-0.010}$}} & {{\small $0.817\pm 0.012$}} &  {{\small $0.866^{+0.026}_{-0.028}$}}                                 
\\\hline
{\small$\sigma_{12}$}  & {{\small$0.792\pm 0.009$}}  & {{\small$0.802^{+0.012}_{-0.011}$}} &  {{\small$0.843^{+0.020}_{-0.022}$}}              
\\\hline
{\small$S_{12}$}  & {{\small$0.794^{+0.011}_{-0.009}$}} & {{\small$0.809^{+0.012}_{-0.015}$}}  &   {{\small$0.855^{+0.023}_{-0.026}$}}  
\\\hline
$r_d$ [Mpc]& {\small$147.26^{+0.20}_{-0.23}$} & {{\small$147.02^{+0.26}_{-0.24}$}} & {{\small$145.06^{+0.97}_{-0.89}$}}       
\\\hline                          
{\small $w$} & {\small$-1$}  & {{\small$-1.063^{+0.033}_{-0.028}$}}  &  {{\small$-1.048^{+0.037}_{-0.034}$}}     

\\\hline
\end{tabular}
\caption{Comparison of the fitting results obtained for the $\Lambda$CDM, $w$CDM and the binned $\rho_{de}(z)$, using the CMBpol+SNIa+$M$+BAO data set.}
\label{table5}
\end{table*}

%%%%%%%%%%%%%%%%%%%%%%%%%%%%%%%%%%
%%%%%%%%%%%%%%%%%%%%%%%%%%%%%%%%%%

EDE can only have a larger impact on the cosmological tensions if $\Omega_{de}(z)$ takes more flexible shapes that allow to respect the very strict constraints found around the CMB decoupling time ($\Omega_{de}(z_{dec})\lesssim 0.4\%$ at $2\sigma$ c.l.), while still leading to a significant EDE fraction in other epochs of the cosmic expansion. The strong bound on EDE for redshift $200<z<1000$ ($\Omega_{de}(z)\lesssim 1\%$ at $2\sigma$ c.l.) is rather impressive since it is entirely based on the different clustering properties of dark energy and dark matter. 

In general, when the SH0ES prior is not included in the fitting analysis there is no significant shift in the value of the Hubble parameter when compared to the $\Lambda$CDM, although the uncertainties clearly grow by a factor $2-3$. This allows to decrease the $H_0$ tension to the $2.66\sigma$ c.l. under the minimal CMBpol+SNIa data set. The addition of the CMB lensing has a very mild effect, increasing the tension up to the $2.73\sigma$ level. 

When the BAO data and the SH0ES prior are also considered, the $H_0$ tension is reduced to the $\sim 2\sigma$ level by increasing the EDE fraction at the radiation-dominated epoch. A $2\sigma$ preference for a non-null EDE density in that epoch is obtained, $\Omega_{de}(z)\lesssim 4\%$ at $95\%$ c.l. Our reconstructed $\Omega_{de}(z)$ allows for non-peaked shapes, in contrast to what one finds e.g. in models based on ultra-light axions \cite{Poulin:2018dzj,Poulin:2018cxd}. The model needs, though, values of the current dark matter density much larger than the ones typically encountered in the concordance model. This is to lower the early integrated Sachs-Wolfe effect down, which is enhanced by the presence of EDE in the pre-recombination epoch. This automatically leads to an increase of $\sigma_8$ and $S_8$ that worsens the LSS tension, which lies now at the $3-3.5\sigma$ c.l. In terms of $\sigma_{12}$, the tension is somewhat lower, but still remains at the $\sim 2.7\sigma$ level, as when the $M$ prior is not used in the analysis. It is also to be noted that the $H_0$ tension still stays at the $3\sigma$ level when formulated in terms of the absolute magnitude of SNIa $M$.

Finally, when we use the most complete data set, taking also the weak lensing data into account, we find that the model can keep the $H_0$ and $S_8$ tensions at the $\sim 2\sigma$ c.l., thanks also to a slight increase of the EDE fraction in the matter-dominated epoch, although, again, the $M$ tension remains at $\gtrsim 3\sigma$. 

In view of our results, it seems unlikely that EDE alone can provide a satisfactory resolution of the cosmological tensions. Whether the latter have or not a physical origin, or whether their statistical significance is as high as claimed by some sectors of the cosmological community, is still a matter of discussion and is certainly not a closed subject. Here we conclude that, in any case, if the data sets employed in this study do not suffer from any important systematic errors, uncoupled EDE is not able to relieve completely the tensions. Under our full data set CMBpolens+SNIa+$M$+BAO+WL they remain at $2-3\sigma$ c.l. 

Nothing prevents, though, the solution to the cosmological tensions to be multi-sided, rather than due to a single new physics component. EDE could still play a significant role in this story. Some interesting directions to explore in the future are: (i) a possible coupling of EDE to dark matter, see e.g. \cite{Karwal:2021vpk}; (ii) the impact of more complicated behavior of the EDE sound speed, parameterizing its time dependence, or performing a tomographic analysis similar to the one we have carried out in this work for the EDE density; and (iii) the potential degeneracy between EDE and other cosmological parameters, as the neutrino masses, which could in principle help to soften the $S_8$ tension while keeping the needed amount of non-relativistic matter at the CMB decoupling time. The latter would be needed, together with a higher EDE fraction in the radiation-dominated epoch, in order to increase the value of the Hubble parameter. We leave these investigations for a future work.    

%%%%%%%%%%%%%%%%%%%%%%%%%%%%%%%%%%%%%%%%%%%%%%%%%%%%
%%%%%%%%%%%%%%%%%%%%%%%%%%%%%%%%%%%%%%%%%%%%%%%%%%%%
%%%%%%%%%%%%%%%%%%%%%%%%%%%%%%%%%%%%%%%%%%%%%%%%%%%%

\vspace{0.25cm}

{\bf Acknowledgements}
\newline
\newline
\noindent
AGV is funded by the Deutsche Forschungsgemeinschaft (DFG) - Project number 415335479.

%%%%%%%%%%%%%%%%%%%%%%%%%%%%%%%%%%%%%%%%%%%
%%%%%%%%%%%%%%%%%%%%%%%%%%%%%%%%%%%%%%%%%%%
%%%%%%%%%%%%%%%%%%%%%%%%%%%%%%%%%%%%%%%%%%%

\appendix

\section{EDEp and quintessence}\label{sec:appendixA}

The EDEp parametrization described in Sec. \ref{sec:pEDE} and analyzed in Sec. \ref{sec:resultsP} is based on the formula \eqref{eq:rhoDE} for the DE density and the covariant self-conservation of the dark energy component, which leads to the expression of its pressure, \eqref{eq:pDE}. We have treated the DE as a perfect fluid. However, for all the epochs of the cosmic expansion in which the effective EoS parameter of the dark energy fluid is larger than -1 it is also possible to formulate the model in terms of a quintessence scalar field $\phi$. In particular, during the RDE and MDE. Let us see how. The energy density and pressure associated to $\phi$ read, respectively, 

\begin{equation}
\rho_\phi = \frac{\dot{\phi}^2}{2}+V\qquad ;\qquad p_\phi = \frac{\dot{\phi}^2}{2}-V\,.
\end{equation}
Hence, we can write $\dot{\phi}$ and the potential as follows, 

\begin{equation}\label{eq:EqPhi}
\dot{\phi}=\sqrt{\rho_\phi+p_\phi}\longrightarrow  d\phi=\frac{da}{a}\sqrt{\frac{3}{8\pi G}}\sqrt{\frac{\rho_\phi(a)+p_\phi(a)}{\rho_m(a)+\rho_r(a)+\rho_\phi(a)}} \,, 
\end{equation}
\begin{equation}\label{eq:potential}
V(a) = \frac{1}{2}\left[\rho_\phi(a)-p_\phi(a)\right]\,,
\end{equation}
identifying the DE perfect fluid density and pressure with the scalar field ones, i.e. doing $\rho_\phi=\rho_{de}$ and $p_\phi=p_{de}$. We can plug now \eqref{eq:rhoDE} and \eqref{eq:pDE} into \eqref{eq:EqPhi} and \eqref{eq:potential}, find $\phi(a)$ from the former and invert it to get $a(\phi)$. This result can be then introduced in \eqref{eq:potential} to finally obtain the shape of the potential needed to reproduce the phenomenology of our EDEp parametrization. For the RDE and MDE the following formula applies,

\begin{equation}\label{eq:integralPhi}
\phi(a)=\phi(a_{ini})+\frac{2}{\kappa}\sqrt{\frac{\chi_1}{1+\chi_1}}\int_{a_{ini}}^{a}\frac{d\bar{a}}{\bar{a}}\left[\frac{1+\frac{3\chi_2}{4\chi_1}\frac{\bar{a}}{a_{eq}}}{1+\left(\frac{1+\chi_2}{1+\chi_1}\right)\frac{\bar{a}}{a_{eq}}}\right]^{1/2}\,,
\end{equation} 
with $\kappa\equiv\sqrt{8\pi G}$, $a_{eq}=a(t_{eq})$ the scale factor associated to the matter-radiation equality time, and $a_{ini}$ the scale factor at some moment deep in the RDE. $\phi_{ini}$ can be safely set to zero, since the quantity that is physically relevant here is $\phi-\phi_{ini}$. Unfortunately, the integral appearing in \eqref{eq:integralPhi} cannot be solved analytically, so we cannot compute the analytical formula of $V(\phi)$. Nevertheless, it is sufficient for our purposes to show that deep enough in the RDE and MDE the potential takes a simple exponential form.

For instance, if $\chi_1\ne 0$ and for sufficiently early times, in the RDE we obtain, 

\begin{equation}
\phi(a) = \frac{2}{\kappa}\sqrt{\frac{\chi_1}{1+\chi_1}}\ln(a/a_{ini})  \,, 
\end{equation}
The potential can be written as follows,

\begin{equation}\label{eq:potentialRD}
V_{\rm RD}(\phi) = \frac{\chi_1}{3}\rho_r(a_{ini})\exp{\left[-2\kappa\phi\sqrt{1+\frac{1}{\chi_1}}\right]}\,,
\end{equation}
whereas in the MDE, for values of the scale factor $a\gg \chi_1 a_{eq}/\chi_2$, we find 

\begin{equation}\label{eq:potentialMD}
V_{\rm MD}(\phi) = \frac{\chi_2}{2}\rho_m(a_{*})\exp{\left[-\kappa\sqrt{3\left(1+\frac{1}{\chi_2}\right)}(\phi-\phi_*)\right]}
\end{equation}
with $a_*$ the scale factor associated to some time in the MDE. Now we can use the scaling formula $\Omega_\phi=\frac{3}{\lambda^2}(1+\bar{w})$ \cite{Copeland:1997et} that applies for quintessence with an exponential potential $V=V_0e^{-\sqrt{8\pi G}\lambda \phi}$, where $\bar{w}$ is the EoS of the dominant component in the Universe. From \eqref{eq:potentialRD} and \eqref{eq:potentialMD} we obtain during the RDE $\lambda^2_{\rm RD}=4(1+\chi_1)/\chi_1=4/\Omega^{\rm RD}_{ede}$, and in the MDE $\lambda^2_{\rm MD}=3(1+\chi_2)/\chi_2=3/\Omega^{\rm MD}_{ede}$. This automatically leads to the formulas \eqref{eq:fractions} for the EDE fractions. This is sufficient to show that there exists a quintessence potential able to  produce the same plateaus in the RDE and MDE that we encounter in the context of the simple EDEp parameterization \eqref{eq:rhoDE}-\eqref{eq:pDE}. This potential reduces to \eqref{eq:potentialRD} and \eqref{eq:potentialMD} in the epochs of interest. The number of additional parameters that we have in this scalar field formulation coincides with the one in the EDEp. Obviously, a full quintessence formulation of the EDEp parametrization in the last stages of the cosmic expansion, during the DE-dominated period, is only possible if $w>-1$ in \eqref{eq:rhoDE}.

%%%%%%%%%%%%%%%%%%%%%%%%%%%%%%%%%%%%%%%%%%%
%%%%%%%%%%%%%%%%%%%%%%%%%%%%%%%%%%%%%%%%%%%

\bibliographystyle{apsrev4-1}
\bibliography{EDE}

\end{document}